\documentclass[showpacs,preprint]{revtex4}
\usepackage{graphicx}
\usepackage{dcolumn}
\usepackage{amsmath}
\usepackage{ amssymb }
\usepackage[colorlinks=true, citecolor=red, urlcolor = red, linkcolor= blue, bookmarks=true]{hyperref}

\begin{document}
\title{Five dimensional rotating regular black holes and shadow}
\author{Fazlay Ahmed$^{a}$} \email{fazleyamuphysics@gmail.com}
\author{Dharm Veer Singh $^{a,\,b}$} \email{veerdsingh@gmail.com}

\author{Sushant G. Ghosh$^{\;a, \;c}$}\email{sghosh2@jmi.ac.in}
\affiliation{$^a$Center for Theoretical Physics,
 Jamia Millia Islamia,  New Delhi 110025,
 India},
\affiliation{$^b$ Depaertment of Physics, Institute of applied Science and Humanities, GLA University, Mathura 281406, Uttar Pradesh, India}

\affiliation{$^c$Astrophysics and Cosmology Research Unit,
 School of Mathematics, Statistics and Computer Science,
 University of KwaZulu-Natal, Private Bag X54001,
 Durban 4000, South Africa}

\begin{abstract}
We present an exact five-dimensional ($5D$) rotating regular black hole metric, with a deviation parameter $k\geq 0$, that interpolates between the $5D$ Kerr black hole ($k=0$) and $5D$ Kerr-Newman ($r \gg k$). This $5D$ rotating regular black hole is an exact solution of general relativity coupled to nonlinear electrodynamics. Interestingly, for a given value of parameter $k$ there exits a critical angular momentum $a=a_E$ which corresponds to extremal rotating regular black hole with degenerate horizons, while for $a<a_E$, one has non-extremal rotating regular black hole with outer and inner horizons. Owing to the correction factor ($e^{-k/r^2}$), which is motivated by the quantum arguments, the ergoregion and black hole shadow are modified.   
\end{abstract}

\maketitle

\section{Introduction}\label{intro}

The general theory of relativity \cite{1} has been subjected to numerous experimental tests starting from astrophysical observations and confirmed experiments. It turns out that experimental results are well in agreement with theoretical predictions of this theory \cite{2}. However, the general relativity predicts that the gravitational collapse of different matter \cite{Joshi:2012mk} affirm the formation of singularities. The existence of a singularity, by its very definition means spacetime ceases to exist signaling a failure of the physical laws. It is widely believed that the black hole singularities do not exist in Nature, but that they are an artifact of classical general relativity.  However, to obtain a comprehensive model for the final fate of the gravitational collapse, one has to use theories that include quantum effects e.g., resolution of the singularities by quantum gravity models \cite{4}. However, we are far from a definite theory of quantum gravity, so try to understand the inside of a black hole and resolve its singularity is to study classical or semi-classical black holes, with regular, i.e., nonsingular properties which can be also  motivated by quantum arguments. Hence, there has significant attention to  find regular black hole solutions with special matter cores that would substitute the true singularities  \cite{Ansoldi:2008jw}. 

The original idea is due to Sakharov \cite{Sakharov:1966aja} who suggested a de Sitter core with equation of state $P=-\rho$ or $T_{ab}=\Lambda g_{ab}$ to get a regular model without singularities, which could provide a proper description at the final stage of gravitational collapse, replacing the future singularity \cite{Gliner:1966}. The idea indicates that an enormous increase in the spacetime curvature during a gravitational collapse process, which may halt it, if quantum fluctuations dominate the process which puts an upper bound on the value of curvature and force the formation of a central core. This idea leads to first regular black hole model by Bardeen \cite{Bardeen:1968}, according to whom there are horizons, but there is no singularity which is a modification of the Reisnner\(-\)Nordstr$\mathrm{\ddot{o}}$m black holes. This triggered several researchers to find the regular black hole models by considering the gravitational field equations in, coupled to nonlinear electrodynamics \cite{10,dym,abg,AyonBeato:2000zs,12,13}, Dymnikova \cite{dym} proposed a series of regular black hole models a with de Sitter core and gives way in a smooth manner into a Schwarzschild solution. 

 Later, Ayon-Beato and Garcia \cite{abg,AyonBeato:2000zs} invoked nonlinear electrodynamics to generate the Bardeen model as an exact nonlinear magnetic monopole, also suggested regular black holes from nonlinear electric fields \cite{AyonBeato:2000zs} which goes exactly encompasses the Reisnner\(-\)Nordstr$\mathrm{\ddot{o}}$m black holes as a special case. Bronnikov \cite{12} proposed several regular black holes in which the source are the fields, the core is an expanding universe with de Sitter asymptotes and the exterior outer region tends to Schwarzschild black hole. In an important development, the general results related to the topology and causality of these regular solutions also reported \cite{13}, Balart \cite{Balart:2009xr} analyzed quasi-local energy of regular black holes. Whereas the thermodynamics of regular black holes was studied  \cite{15} as well as black hole as particle accelerator \cite{16}. Lemos and Zanchin \cite{Lemos:2011dq} offers an up-to-date classification to discuss the types of regular black holes derived  and presents a new solution with  de Sitter core (see also Ansoldi \cite{Ansoldi:2008jw}, for a review on regular black holes). Subsequently, also there  has been intense activities in the investigation of regular black holes \cite{Hayward:2005gi}, and more recently by \cite{19, Xiang:2013sza,Toshmatov}, but most of these solutions are more or less based on Bardeen's proposal.

However, the rotating black holes are more important in the astrophysical observations, as the black hole spin plays a critical and key role in any astrophysical process. This lead to generalization in finding the Kerr-like regular black holes starting with Bambi and Modesto \cite{Bambi:2013ufa} who constructed rotating regular Bardeen and Hayward black hoes, but the weak energy condition (WEC) is violated, which  was generalized by Neves and Saa \cite{Neves:2014aba} to accommodate a cosmological constant. Further,  generalization of regular solutions to the axially symmetric case \cite{Ghosh:2014pba,Larranaga:2014uca, Toshmatov}, via the Newman-Janis algorithm \cite{Newman:1965tw} and by other similar techniques \cite{27}.

It would be interesting to look at the generalization of regular solutions to higher dimensions, and discuss its properties. The purpose of this paper is to obtain an exact three parameter stationary, axisymmetric metrics that describe five dimensional ($5D$) rotating regular (nonsingular) black holes. The metrics depend on the mass ($M$) and spin ($a$) as well as a free parameter $(k)$ that measure potential deviation from the $5D$ Kerr solution \cite{kerr} and also go over to the $5D$ Kerr-Newman solution, i.e., the generalization of four dimensional regular solutions \cite{Xiang:2013sza,Ghosh:2014pba}.

In this paper, we obtain a $5D$ regular black hole metric for a nonlinear electrodynamics as source in Sec.~\ref{5Drbh} and also give the basic equations of gravity theory. In Sec.~\ref{rrbh}, we study the $5D$ rotating regular black hole and also investigate the structure and location of the horizons along with regularity conditions. Particle motion and shapes of the shadow and the energy emission rate have studied in Sec.~\ref{pmbh}. Finally we conclude in the~ Sec.~\ref{cnbh}.

\section{Five dimensional regular black hole}\label{5Drbh}
We consider the theory of gravity minimally coupled to nonlinear electrodynamics in $5D$ manifold, which comes from the action
\begin{equation}
\label{1} 
\begin{split}
S=\int
d^5x\sqrt{-g}\left[R-{\cal{L}}(F)
\right],
\end{split}
\end{equation}
where $R$ is the Ricci scalar, and ${\cal{L}}(F)$ is an arbitrary function of $F$. The matter is described by the nonlinear electrodynamics Lagrangian ${\cal{L}}(F)$ with $F=F_{\mu \nu}F^{\mu \nu}/4$, where $F_{\mu \nu}$ is associated with gauge $A_{\mu}$ as $F_{\mu \nu}=\partial_{\mu} A_{\nu}-\partial_{\nu} A_{\mu}$. The ${\cal{L}}(F)$ takes the form \cite{Ghosh:2018bxg}
\begin{equation}
{\cal{L}}(F)= 3 F\,e^{-\frac{k}{e}(2eF)^{1/3}},
\label{2}
\end{equation}
where $k>0$ is a parameter. 
  Obviously, when ${\cal{L}}(F)=F$, one gets $5D$ Reisnner\(-\)Nordstr$\mathrm{\ddot{o}}$m like charged black holes. Varying the action (\ref{1}), we obtain the equations of motion \cite{Ghosh:2018bxg,Hendi:2017phi,Panahiyan:2018gzq}
\begin{eqnarray}
&&R_{ab}-\frac{1}{2}g_{ab}R= T_{ab}\equiv2\left[\frac{\partial {\cal{L}}(F)}{\partial F}F_{a c}F_{b}^{c}-g_{a b}{\cal{L}}(F)\right],
\label{egb2}
\end{eqnarray}
\begin{eqnarray}
&& \nabla_{a}\left(\frac{\partial {\cal{L}}(F)}{\partial F}F^{a b}\right)=0\qquad \text{and} \qquad \nabla_{\mu}(* F^{ab})=0,
 \label{egb3}
\end{eqnarray}
where ${\cal{L}}(F)$ is the Lagrangian density of nonlinear electromagnetic field which is a function of  $F$. It turns out that for a $5D$ spherically symmetric spacetimes, the only non-vanishing components of $F_{ab}$ are  $F_{\theta\phi}, F_{\theta\psi}$ and $F_{\phi\psi}$. The Maxwell field for nonlinear electrodynamics in a $5D$ spacetime can be written as \cite{Ghosh:2018bxg}
\begin{equation}
F_{ab}=2\delta^{\theta}_{[a}\delta^{\phi}_{b]} \zeta(r,\theta,\phi),
\label{8}
\end{equation}
where $\zeta$ has been suitably modified for the $5D$. Using Eqs.~(\ref{egb3}) and (\ref{8}), we get 
\begin{equation}
F_{ab}=2\delta^{\theta}_{[a}\delta^{\phi}_{b]}e(r) \sin^2\theta\sin\phi.
\label{9}
\end{equation}
Eq.~(\ref{egb2}) implies $dF=0$, hence we have
\begin{equation}
 e^{\prime}(r)\sin^2\theta\sin\phi dr\wedge d\theta\wedge d\phi\wedge d\psi=0,
\end{equation}
which leads to $e(r)= e$ = constant. Interestingly, the other components of $F_{ab}$ have negligible influence in comparison to $F_{\theta\phi}$ \cite{Ghosh:2018bxg}. Hence the field strength can be simplified 
\begin{equation}
F_{\theta\phi}=\frac{e}{r}\sin\theta,\qquad \text{and} \qquad  F=\frac{e^2}{2r^6}.
\label{10}
\end{equation}
Using Eq.~(\ref{10}) in Eq.~(\ref{2}), we obtain
\begin{equation}
 {\cal{L}}(F)=\frac{3 e^2 }{r^6}e^{-e^2/{M r^2}},
\label{11}
\end{equation}
here, the charge $e$ and mass $M$ are related by the parameter via $e^2=Mk$.
For the $5D$ static spherically symmetric case, using Eq.~(\ref{egb3}), (\ref{9}), (\ref{10}) and (\ref{11}), the energy momentum tensor reads
\begin{equation}
T^t_t=T^r_r=\rho(r)=\frac{3Mk}{r^6}e^{-k/r^2}.
\label{ttr}
\end{equation}
The Bianchi identity $T^{ab}_{;b}=0$ \cite{Rizzo:2006zb}, gives
\begin{eqnarray}
\partial_r T^r_r + \frac{1}{2} g^{tt} [T^r_r-T^t_t] \partial_r g_{tt} + \frac{1}{2} \sum_{i=1}^{3} g^{ii} [T^r_r-T^i_i] \partial_r g_{ii}=0.
\end{eqnarray}
So, the complete set of energy-momentum tensor in $5D$ spacetime is
\begin{eqnarray}
 && T^t_t=T^r_r=\rho(r), \nonumber\\
&& T^{\theta}_{\theta} = T^{\phi}_{\phi} = T^{\psi}_{\psi} =\rho(r)+ r \partial_r \rho(r)/3.
\end{eqnarray}

We are interested in an exact $5D$ spherically symmetric regular black hole solution of Einstein gravity. The general $5D$ spherically symmetric spacetime can be written as
\begin{equation}
ds^2=-f(r)dt^2+\frac{1}{f(r)}dr^2+r^2d\Omega_3^2,
\label{ans11}
\end{equation}
where $d\Omega_3^2=d\theta^2+\sin^2\theta \,(d\phi^2+\sin^2\phi \,d\psi^2)$ is the metric on the $3D$  sphere. Using the metric \textit{ansatz} (\ref{ans11}), the ($r,r$) component of Eq.~(\ref{egb2}) reads
\begin{eqnarray}
2 r(1-f(r))-r^3f'(r)=\frac{2Mk}{r^3}e^{-k/r^2},
\label{eom1}
\end{eqnarray}  
where a prime denotes a derivative with respect to $r$. The Einstein field equation (\ref{eom1}) coupled to nonlinear electrodynamics admits a general solution
\begin{equation}
ds^2=-\left[1-\frac{Me^{-k/r^2}}{r^2}\right]\,dt^2+ \frac{1}{\left[1-\frac{Me^{-k/r^2}}{r^2}\right]} dr^2+r^2\,d\Omega_3^2.
\label{eqn.gb}
\end{equation}

Thus we have an exact $5D$ spherically symmetric solution of Eq.~(\ref{egb2}) with nonlinear electrodynamics as a source (\ref{ttr}). The metric~(\ref{eqn.gb}) is a generalization of $5D$ Schwrazschild-Tengherleni black hole, which encompassed as a special case when $k=0$. We approach regularity properties of solution by observing the behaviour of curvature invariants $R$, $\mathcal{R} =R_{ab} R^{ab}$ and $K=R_{abcd}R^{abcd}$, where $R_{ab}$ and $R_{abcd}$ are the Ricci tensor and Riemann tensor, respectively. The invariants are calculated as

\begin{eqnarray}
&&R= \frac{2 k M e^{-k/r^2}}{r^8} \big(-r^2 + 2 k \big),   \nonumber\\
&&\mathcal{R} = \frac{4 k^2 M^2 e^{-2k/r^2}}{r^{16}}\big(2 k^2  -8 k r^2 + 11 r^4 \big),   \nonumber \\
&&K = \frac{4 M^2 e^{-2k/r^2}}{r^{16}}\big(4 k^4  - 28 k^3 r^{2} + 67 k^2 r^{4}-54 k r^{6}+18 r^{8}\big).
\end{eqnarray}
These invariants are well behaved everywhere for $M \neq 0$, including at $r=0$ and hence the metric~(\ref{eqn.gb}) is regular. However, for the Reissner\(-\)Nordstr$\mathrm{\ddot{o}}$m case ($r\gg k$), they diverge at $r=0$ indicating a scalar polynomial singularity \cite{Wheeler}. In that solution the generalization of the $4D$ spherically symmetric regular black hole \cite{Bambi:2013ufa,Singh:2017qur} to rotating case or Kerr-like solutions obtained. It is demonstrated that the rotating regular solutions can be derived starting from the corresponding exact spherically symmetric solutions by the Newman-Janis  algorithm \cite{Newman:1965tw}. The rotating regular black hole solution includes the Kerr metric as a special case if the deviation parameter $k=0$, and asymptotically ($r \gg k$), it behaves like Kerr-Newman black hole.

\section{ Rotating Regular black hole}\label{rrbh}
The Newman-Janis is an ingenious algorithm to construct a metric for rotating black hole starting from static spherically symmetric solutions. Applying the Newman-Janis algorithm in Eq.~(\ref{eqn.gb}), we obtain the $5D$ rotating regular Kerr-like black hole which in the Boyer-Lindquist coordinates reads
\begin{eqnarray}\label{metric}
d s^2 & = & -\Big[1-\frac{M e^{-k/r^2}}{\Sigma}\Big]dt^2-\frac{2 M e^{-k/r^2}a \sin^2 \theta}{\Sigma}dt d \phi + \frac{\Sigma}{\Delta}d r^2 + \Sigma d \theta^2  \nonumber \\ && +\Big[r^2+a^2 +\frac{ M e^{-k/r^2} a^2 \sin^2 \theta}{\Sigma}\Big]\sin^2 \theta d \phi^2 + r^2 \cos^2 \theta d \psi^2,
\end{eqnarray}
where $\Sigma=r^2+a^2 \cos^2 \theta$ and $\Delta=r^2+a^2- M e^{-k/r^2}$, and three parameters $M$, $a$ and $k$ are assumed to be positive. 
The metric~(\ref{metric}) includes $5D$ Kerr solution as a special case when $k=0$, and Schwarzschild solution when $k=a=0$. The solution for $M=0$, is flat. Obviously, for $a=0$, it goes over to the spherically symmetric metric (\ref{eqn.gb}). This metric in $4D$, has been analysed in \cite{Ghosh:2014pba}.
\begin{figure*}[h]
    \begin{tabular}{c c}
        \includegraphics[scale=0.45]{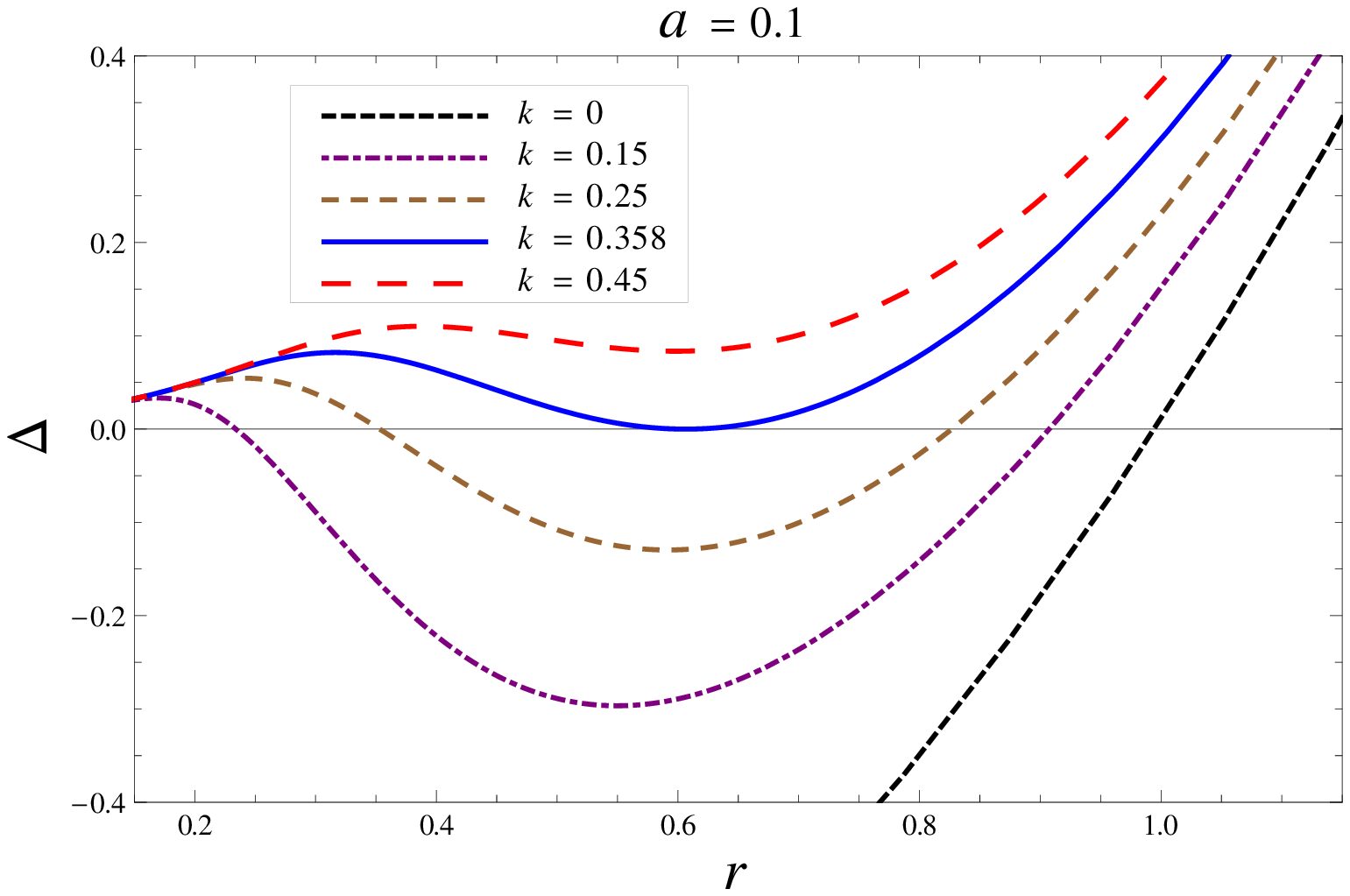}&
        \includegraphics[scale=0.45]{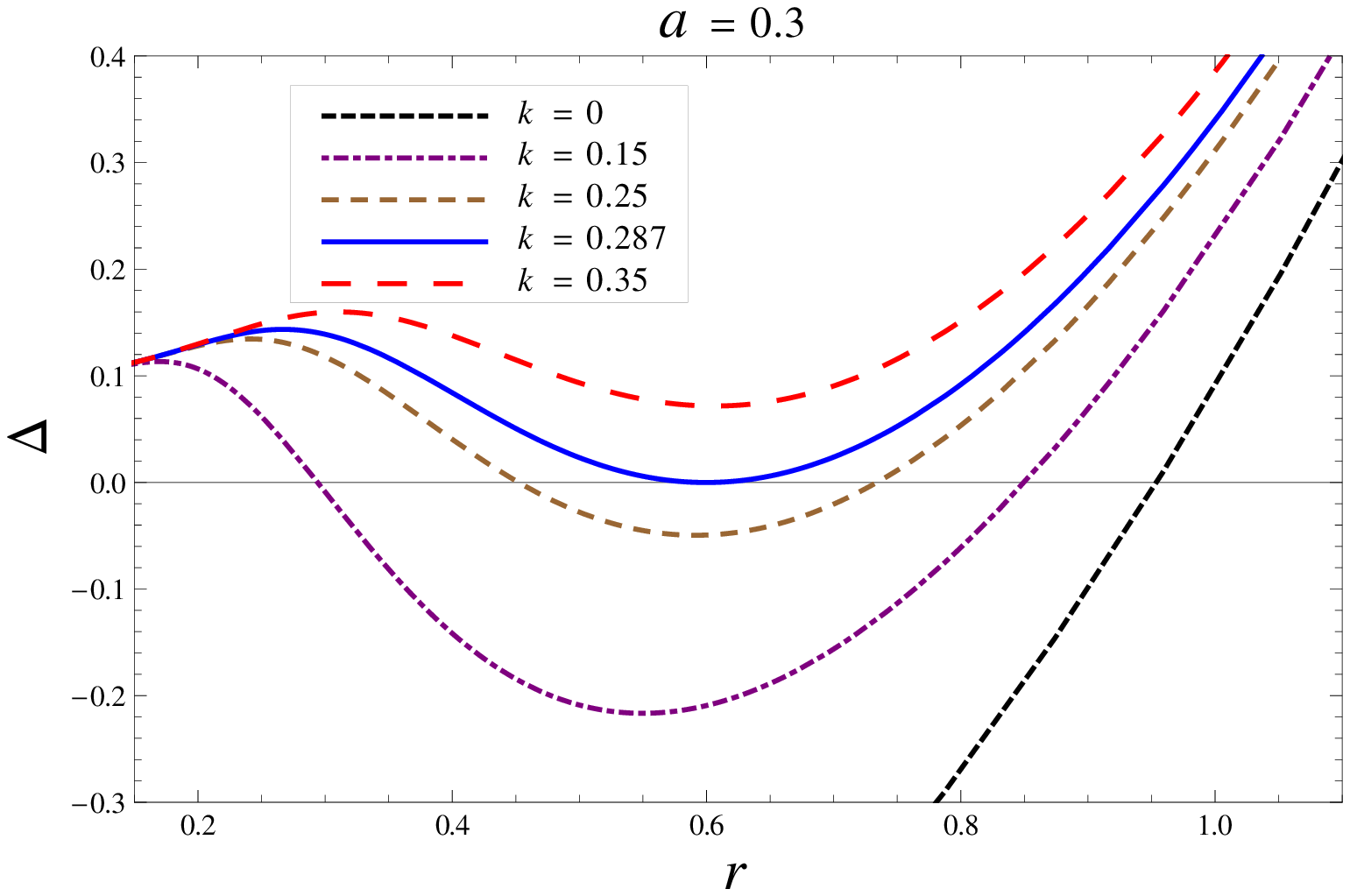}
    \end{tabular}
    \caption{Plots showing the behaviour of $\Delta$ vs radius $r$ for different values of deviation  parameter $k$, the case $k=0$ corresponds to the $5D$ Kerr black hole. }
\label{ehf}
\end{figure*}

\noindent Note that the metric~(\ref{metric}) asymptotically ($r^2 \gg k$) behaves as 
\begin{eqnarray}
g_{tt} &=& 1- \frac{M}{\Sigma}+\frac{e^2}{r^2 \Sigma} + O(e^4/r^4),  \nonumber\\ 
\Delta &=& r^2 + a^2 - M + \frac{e^2}{r^2} + O(e^4/r^4).
\end{eqnarray}


\subsection{Horizon structure and ergoregion}\label{hrzn}
The event horizon is a null surface, determined by $\Delta=0$, implies
\begin{equation}
 r^2+a^2- M e^{-k/r^2}=0,
\label{delta}
\end{equation}
which may have many zeros. The largest root of Eq.~(\ref{delta}), give the location of the black hole's event horizon. When $k=0$, one obtains the $5D$ Kerr black hole of general relativity. Here we discuss the effect of parameter $k$ on the horizons and ergoregions. Clearly the event horizon radii depend on $k$, which is different from the usual Kerr case.

\begin{figure*}
\begin{tabular}{c c c c}
\includegraphics[scale=0.38]{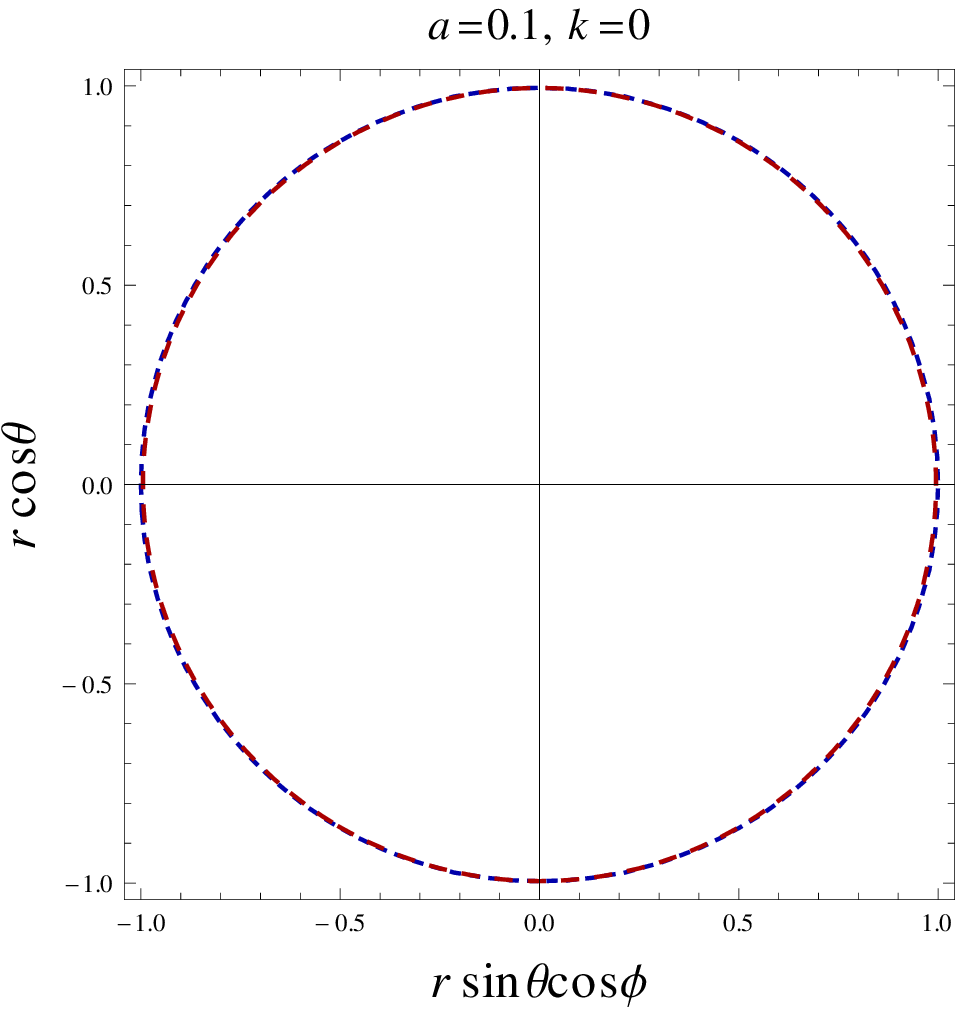}&
\includegraphics[scale=0.38]{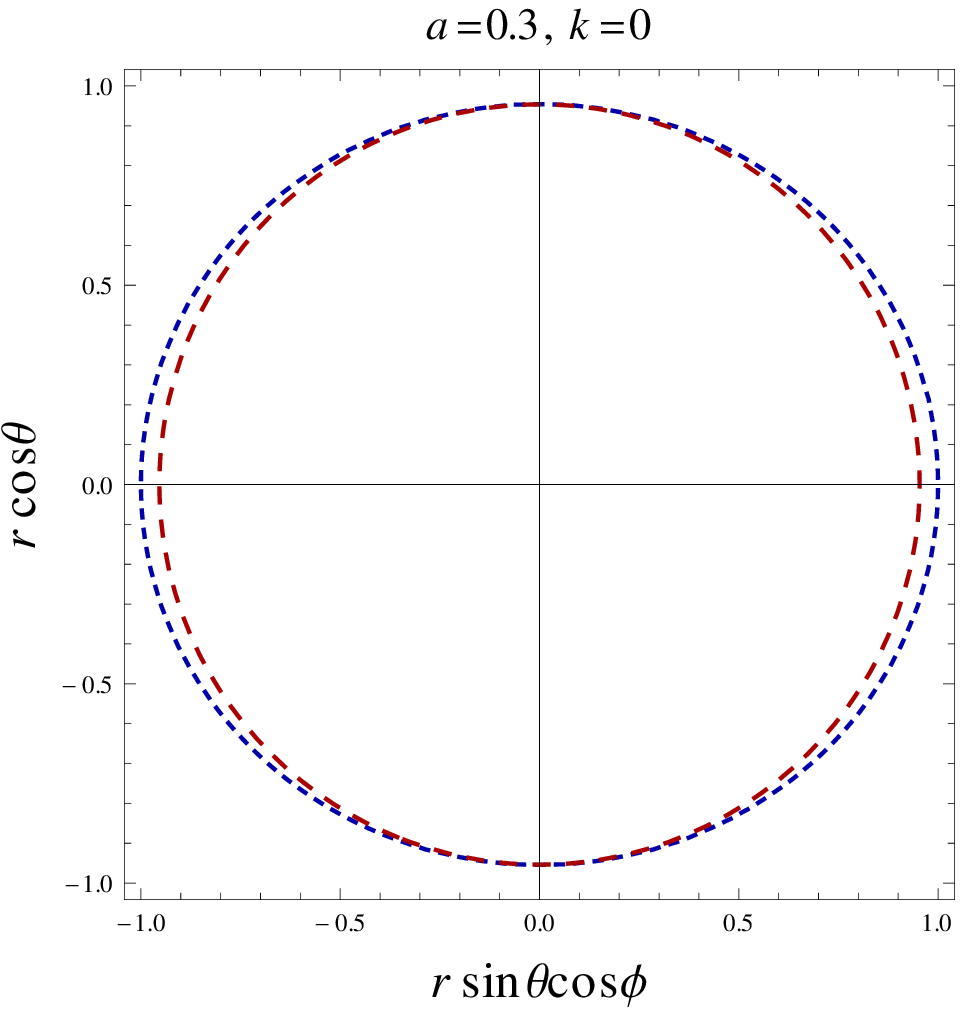}&
\includegraphics[scale=0.38]{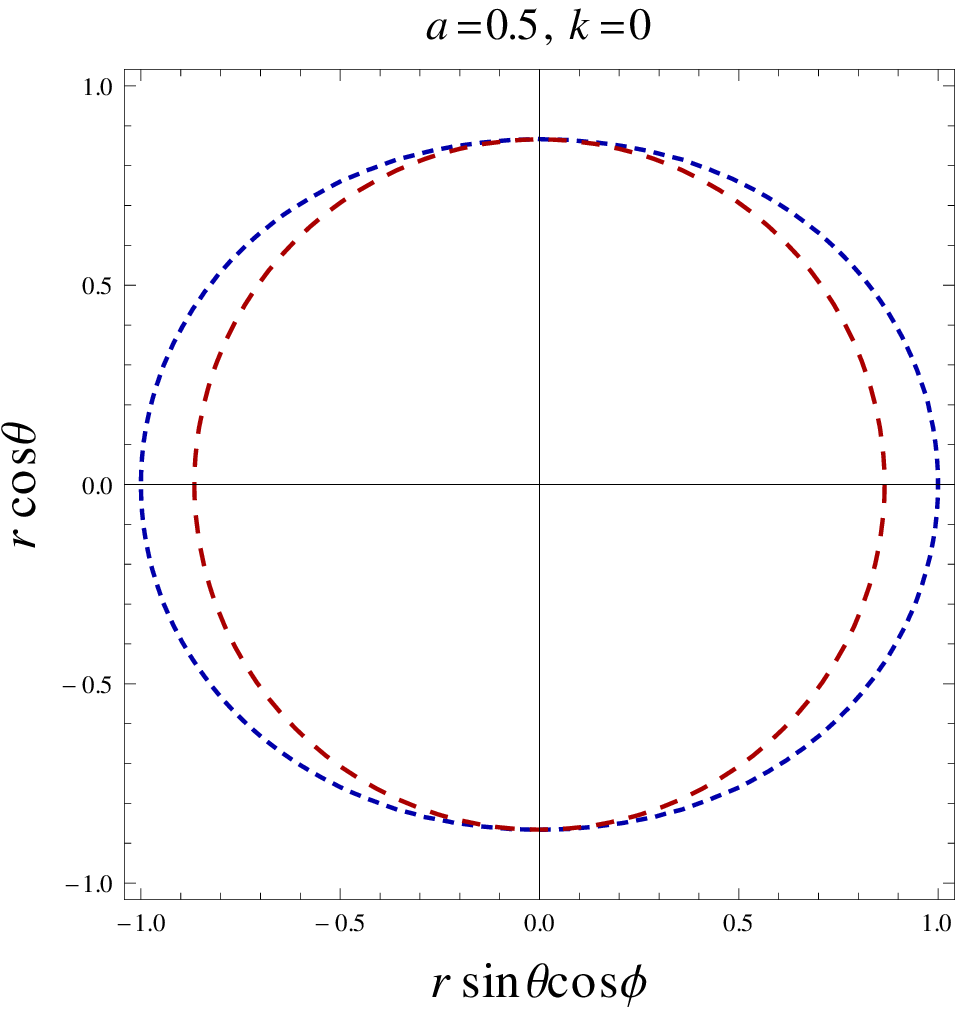}&
\includegraphics[scale=0.38]{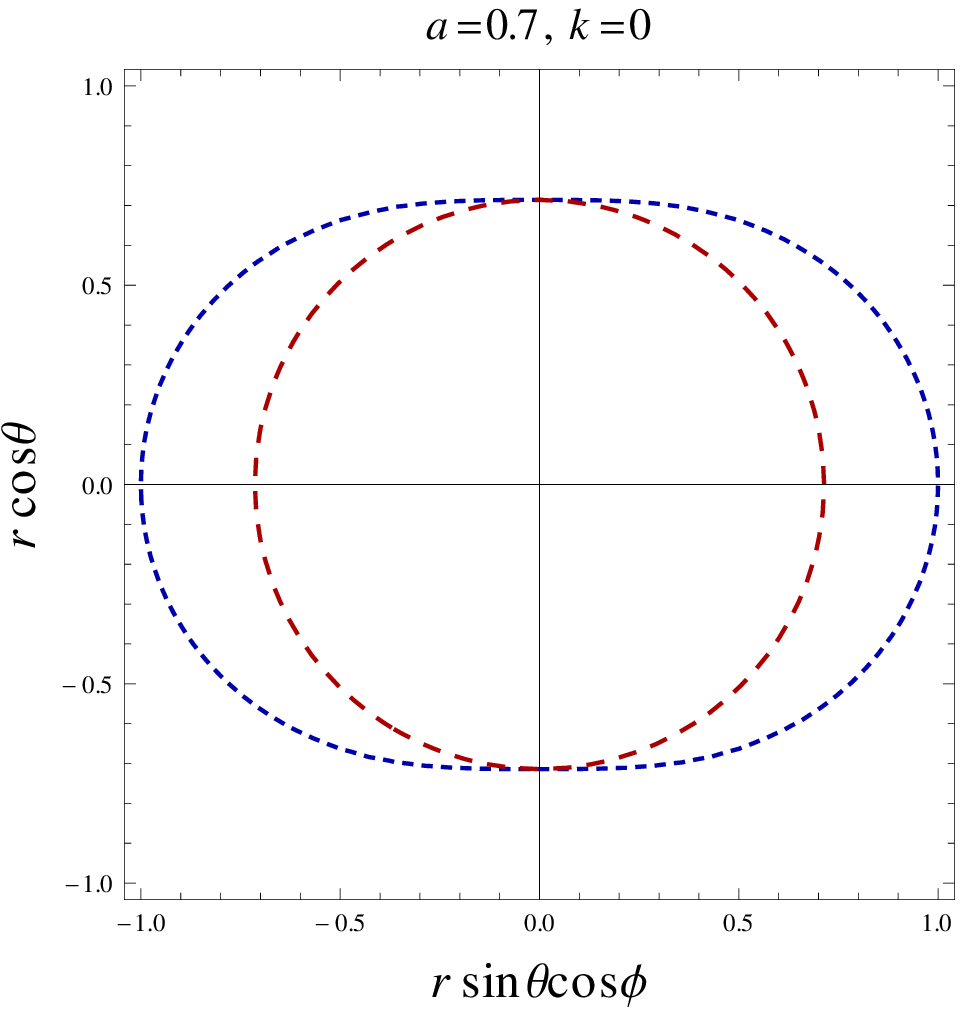}\\
\includegraphics[scale=0.38]{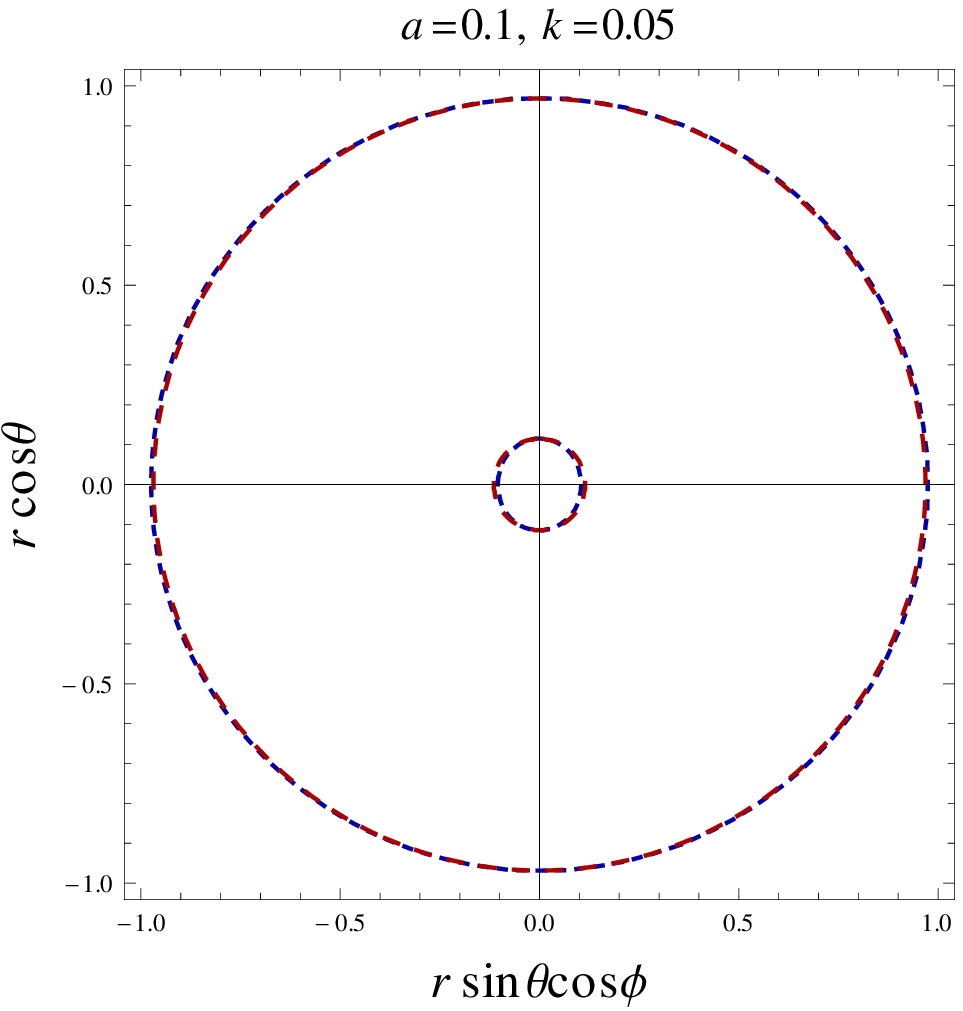}&
\includegraphics[scale=0.38]{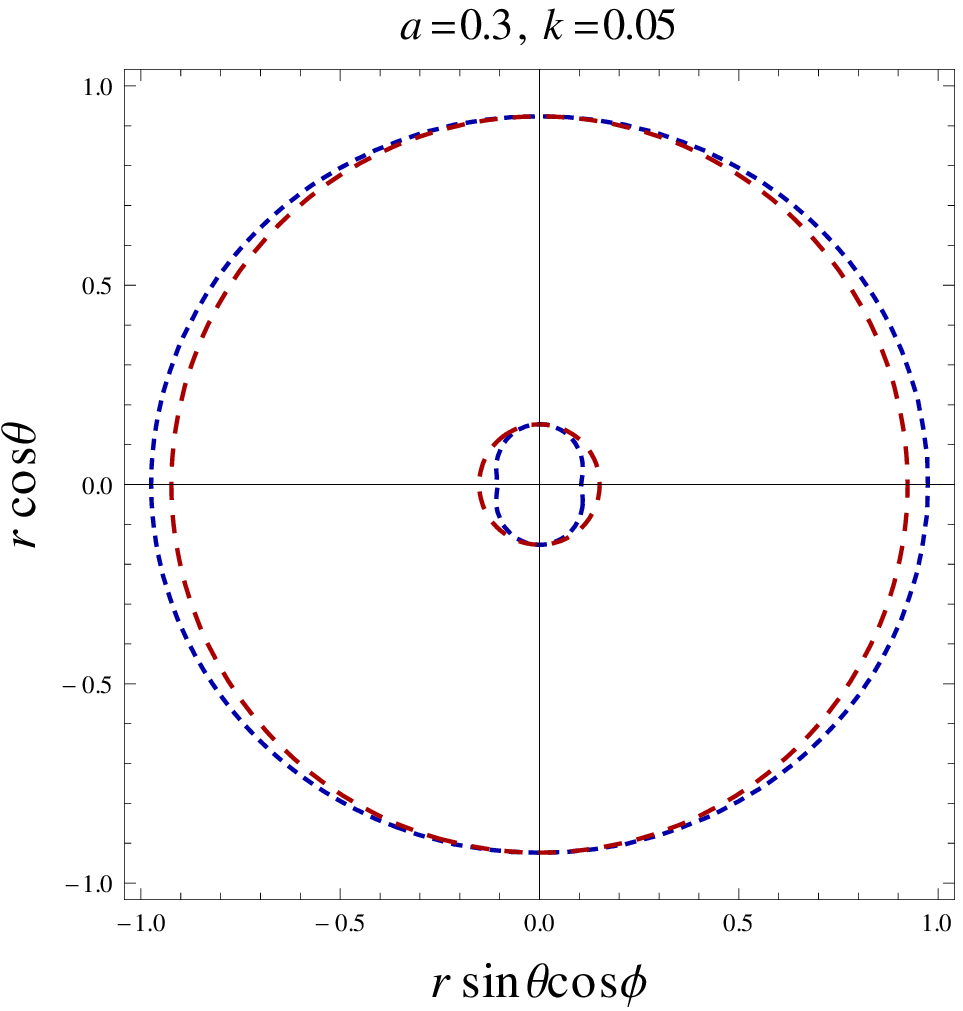}&
\includegraphics[scale=0.38]{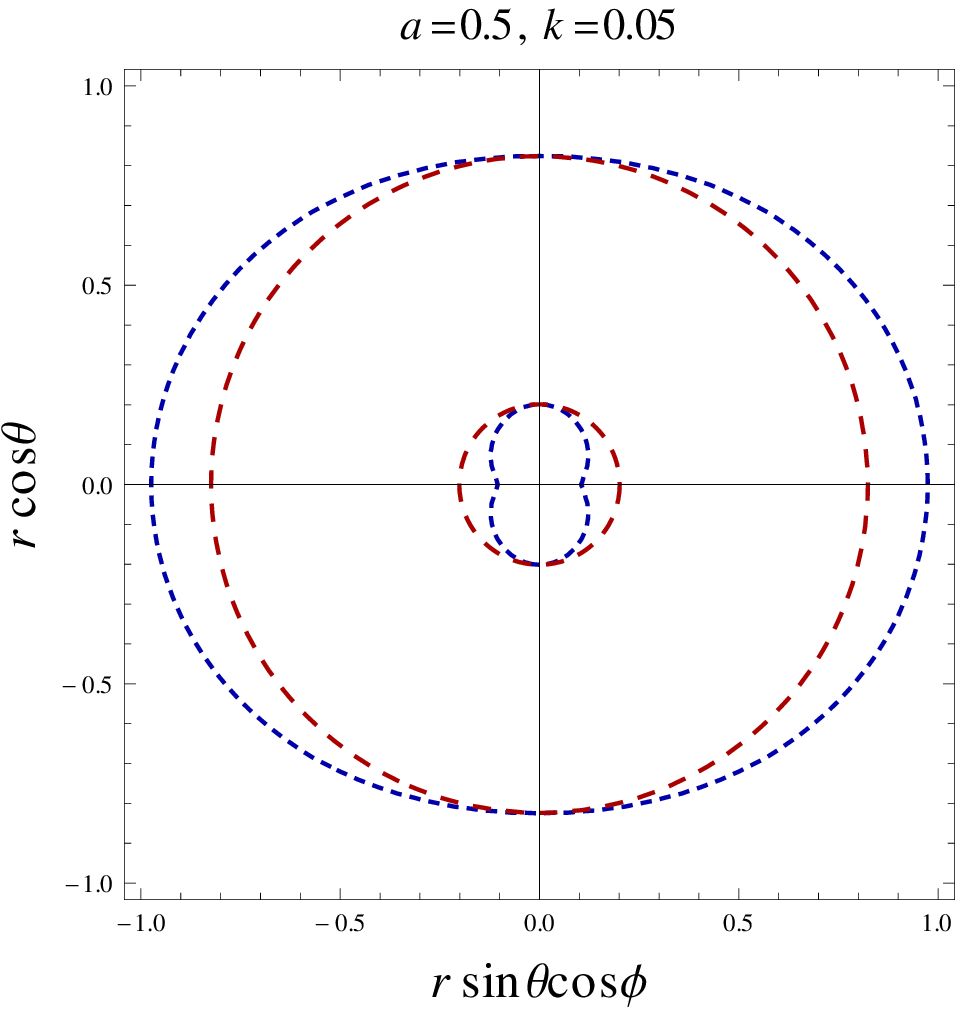}&
\includegraphics[scale=0.38]{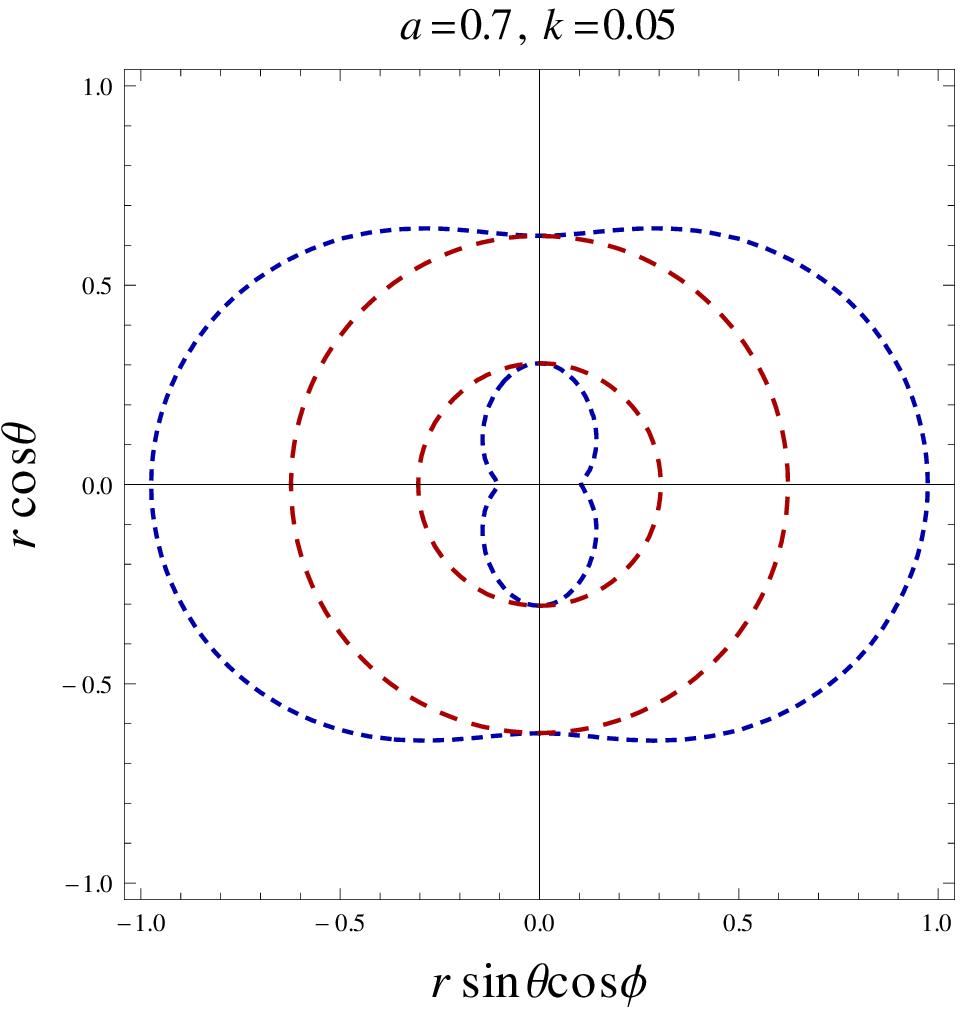}\\
\includegraphics[scale=0.38]{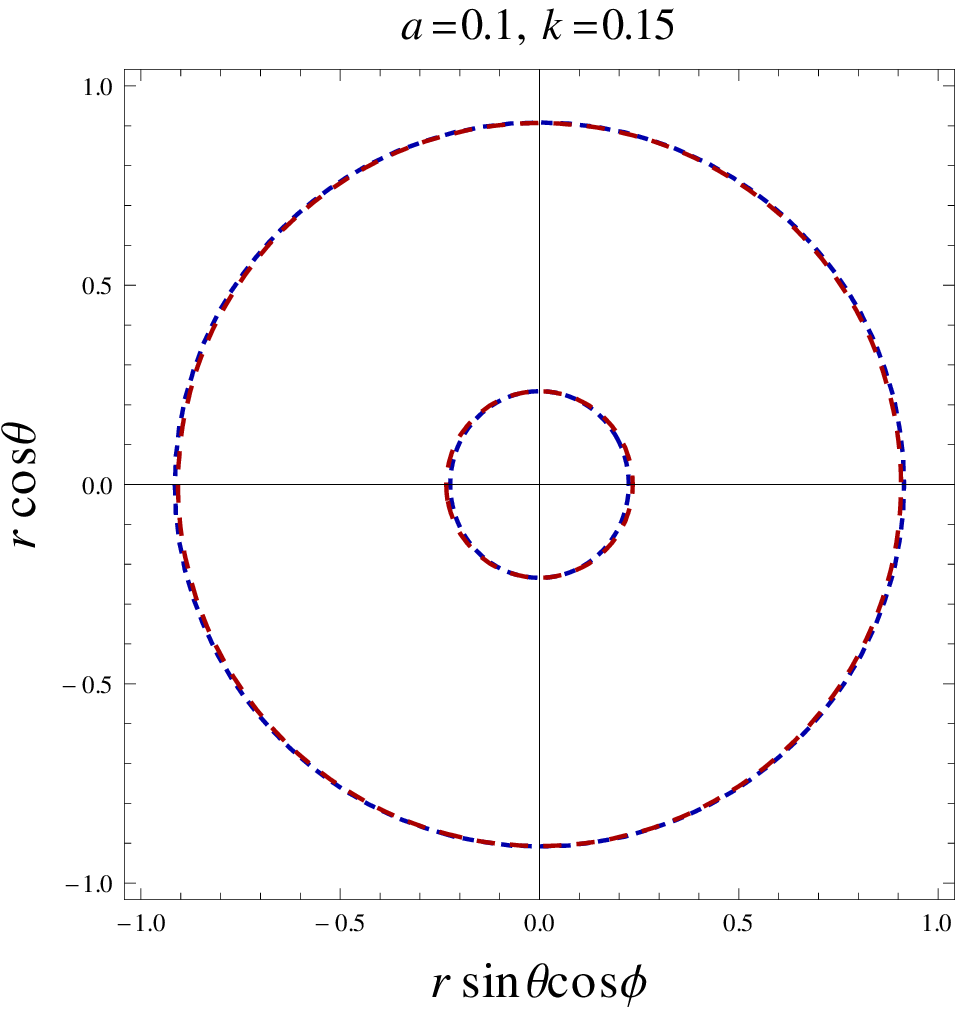}&
\includegraphics[scale=0.38]{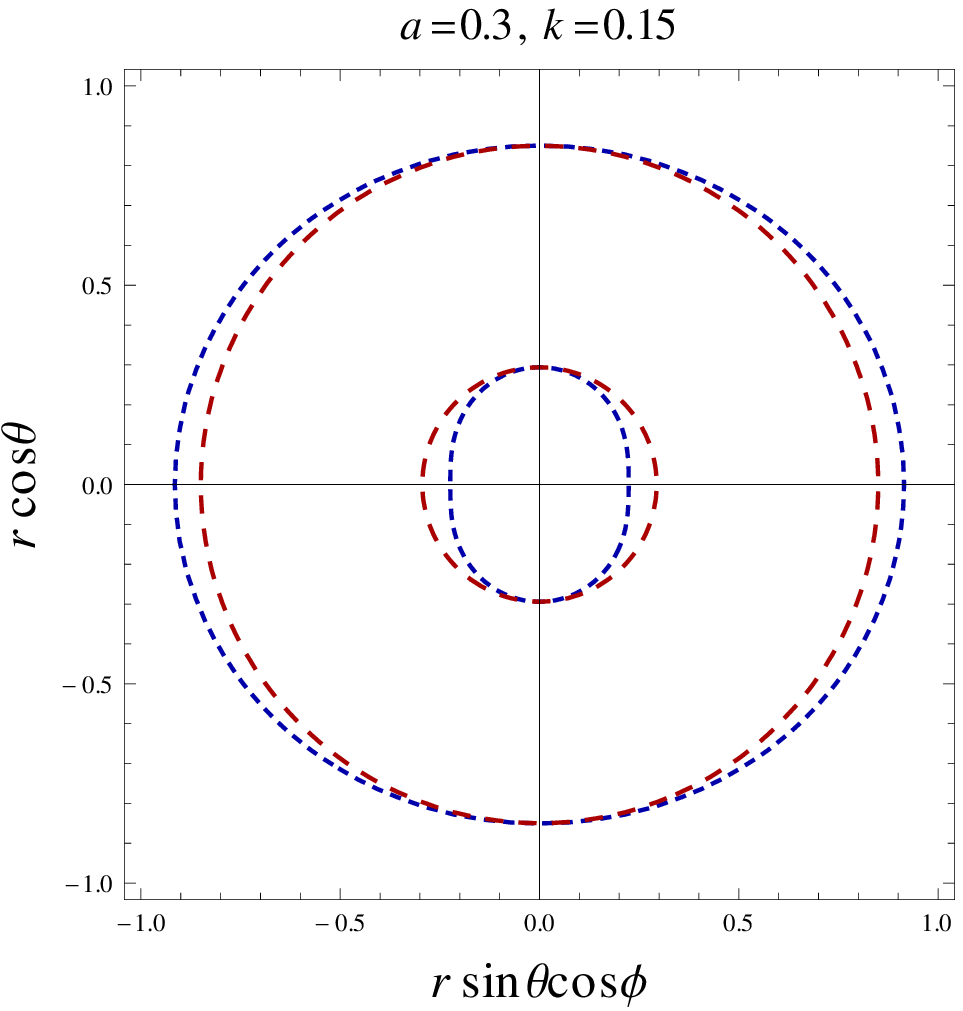}&
\includegraphics[scale=0.38]{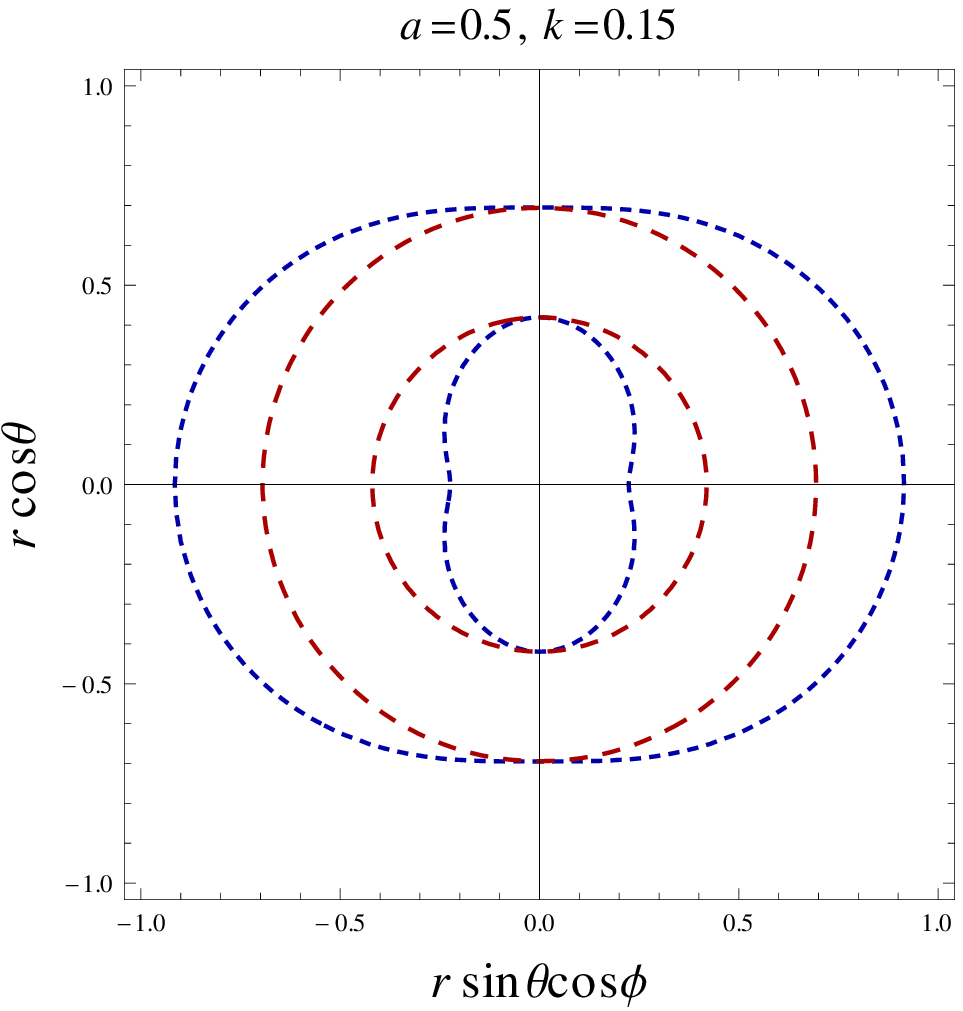}&
\includegraphics[scale=0.38]{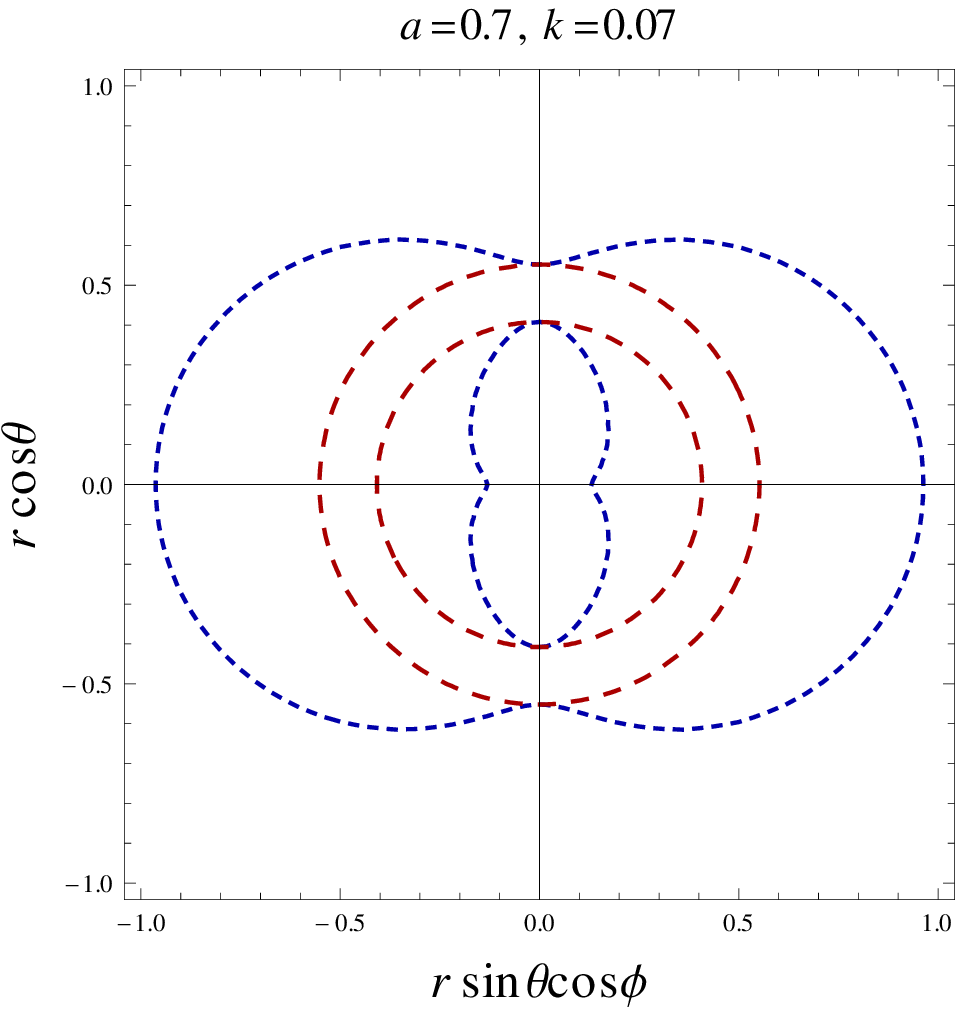}\\
\includegraphics[scale=0.38]{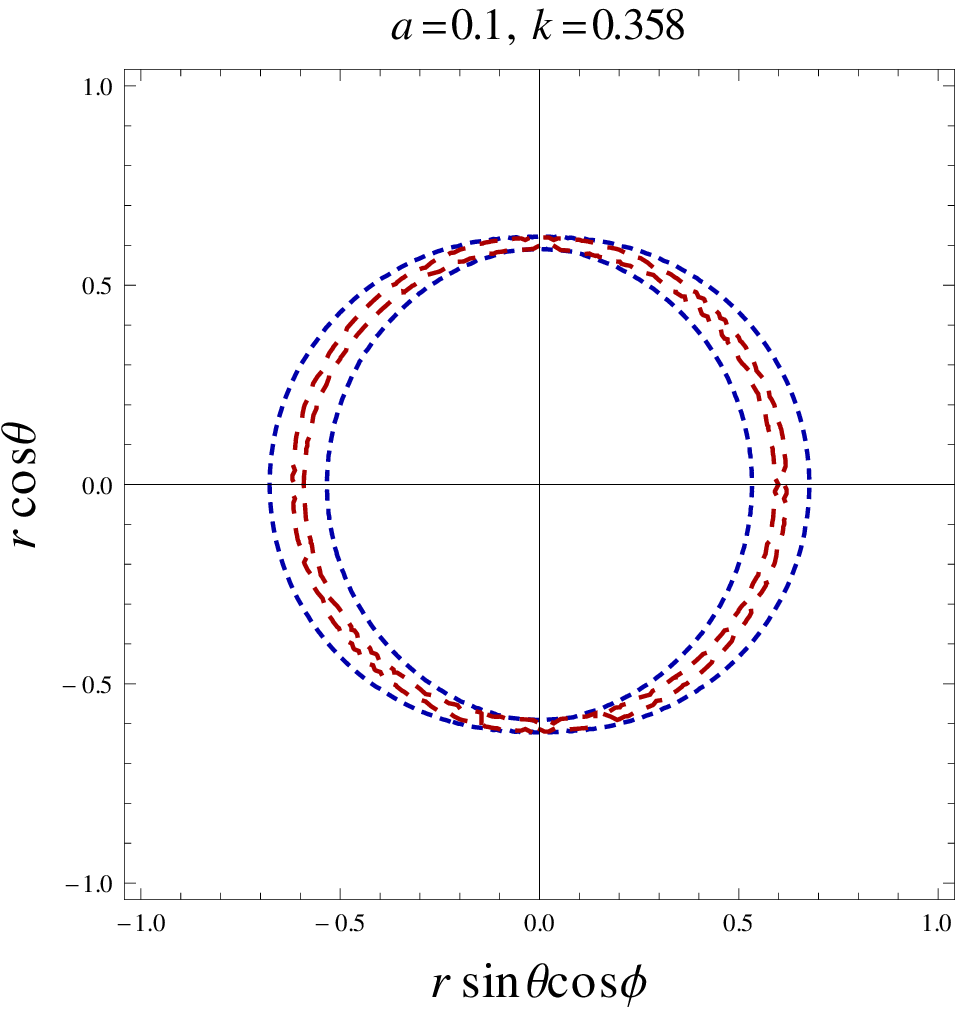}&
\includegraphics[scale=0.38]{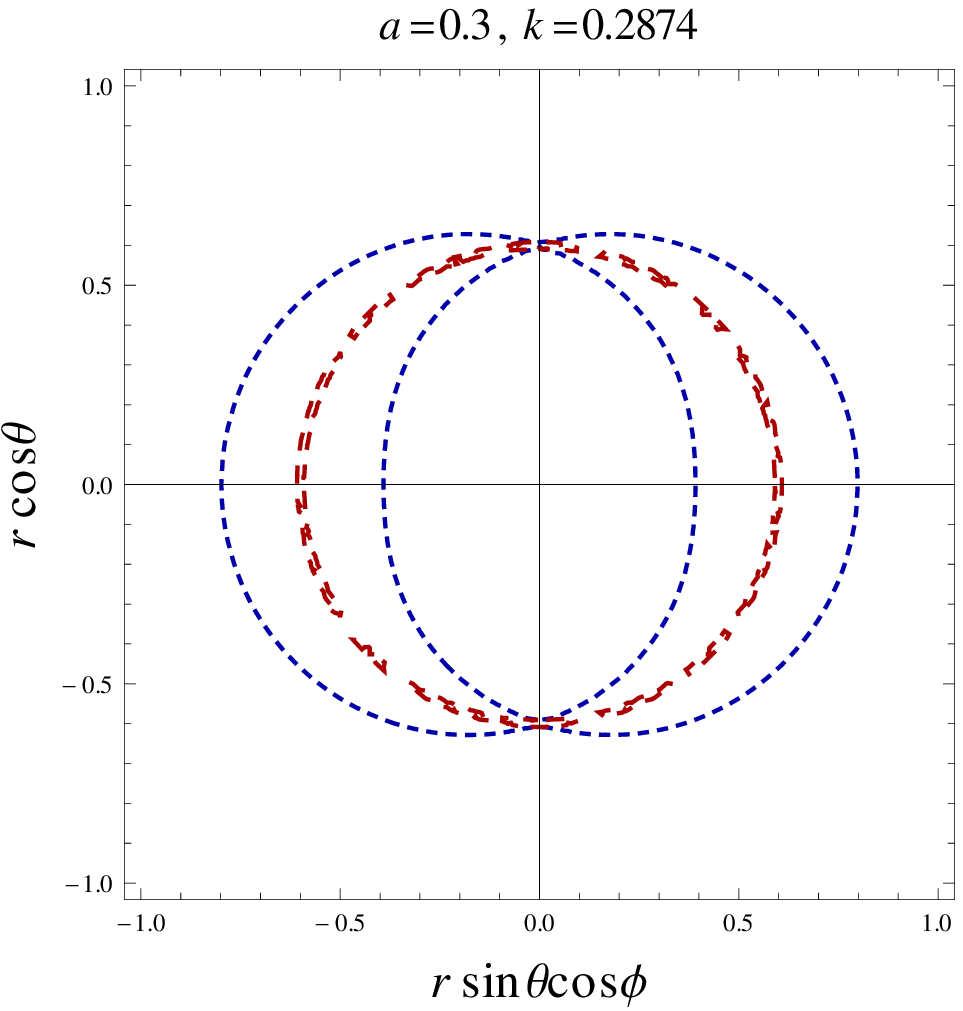}&
\includegraphics[scale=0.38]{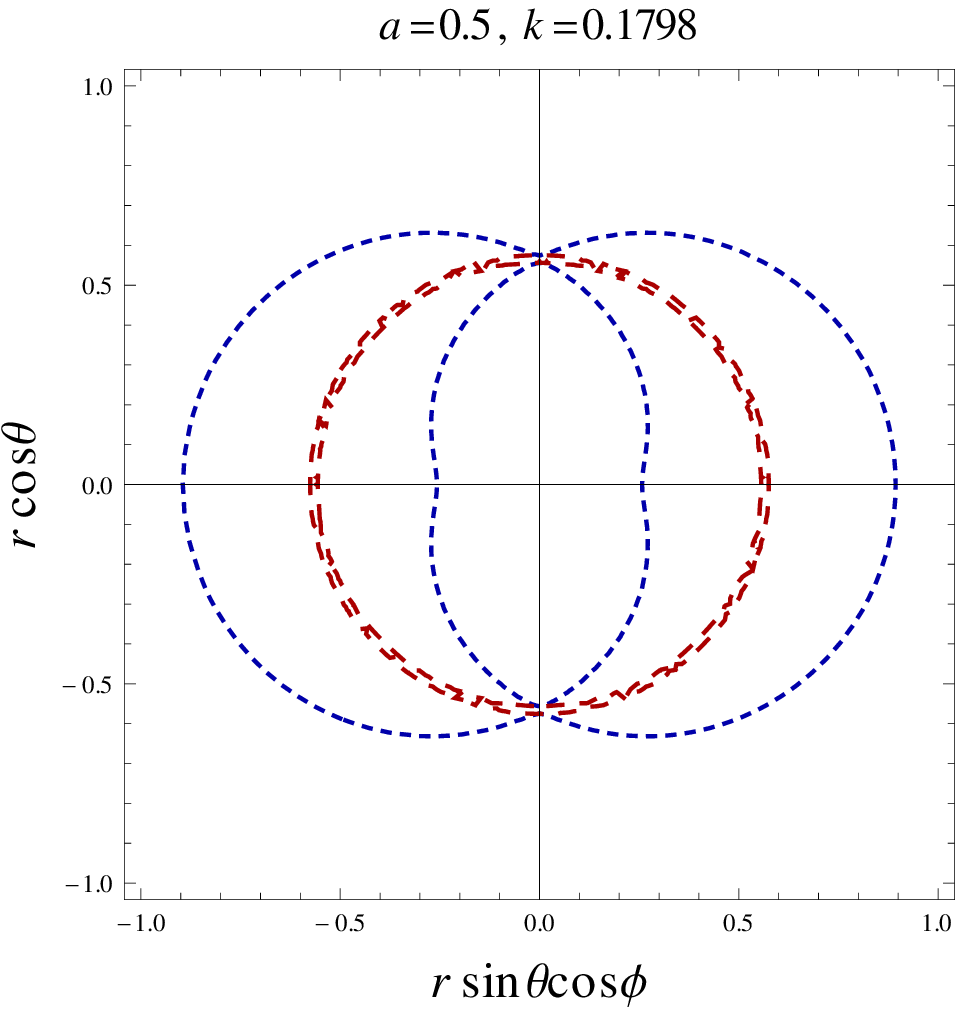}&
\includegraphics[scale=0.38]{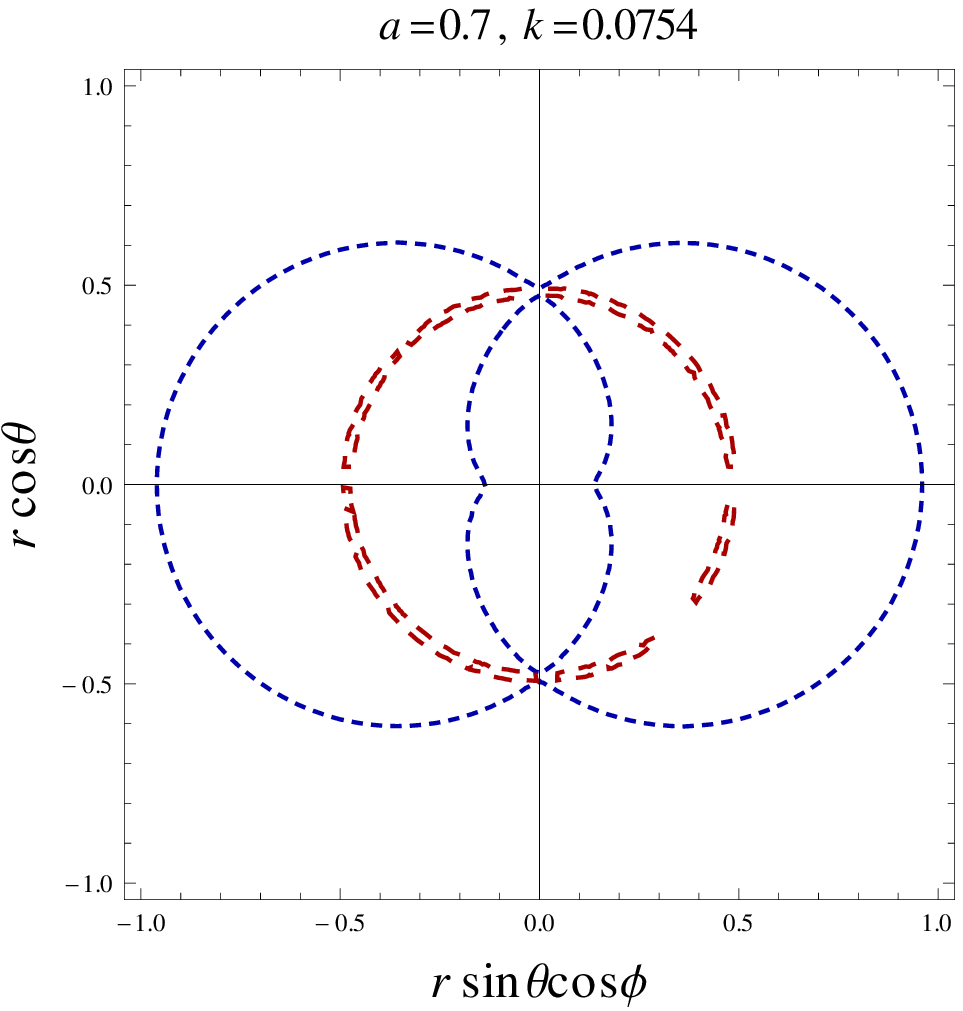}\\
\includegraphics[scale=0.38]{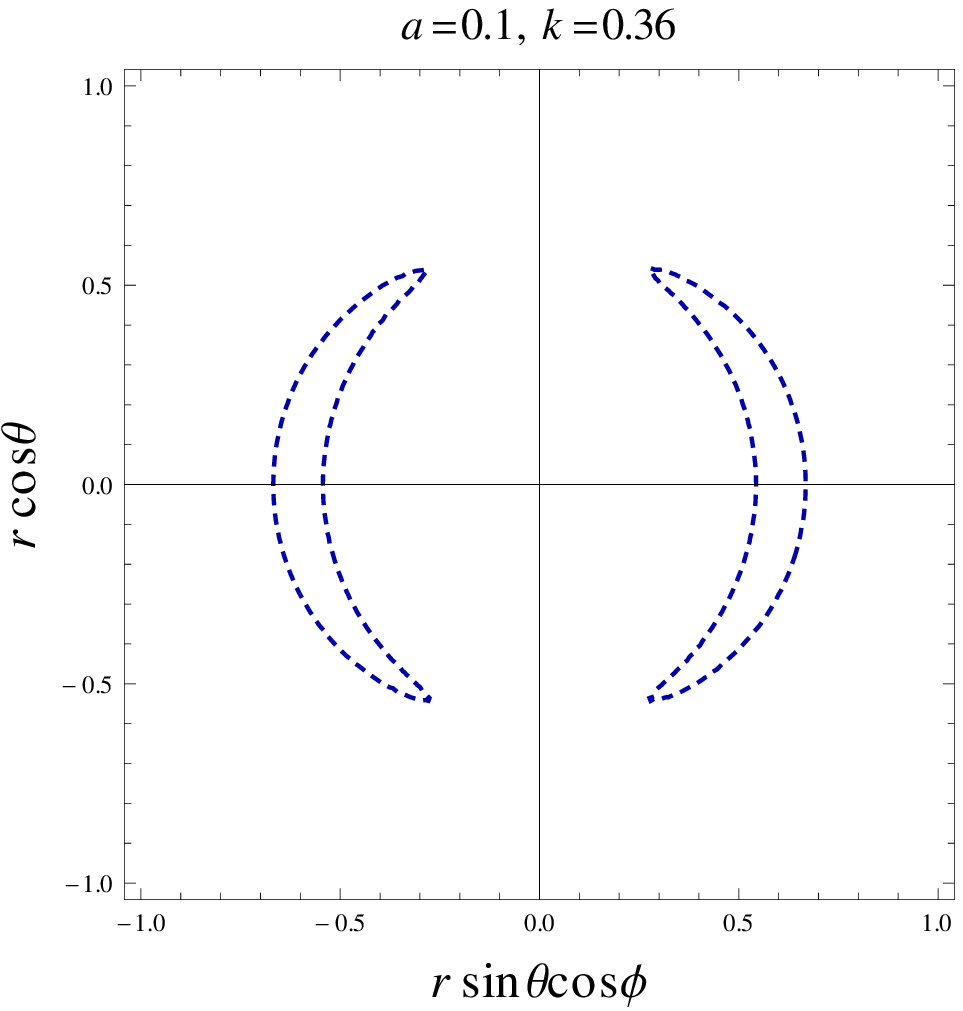}&
\includegraphics[scale=0.38]{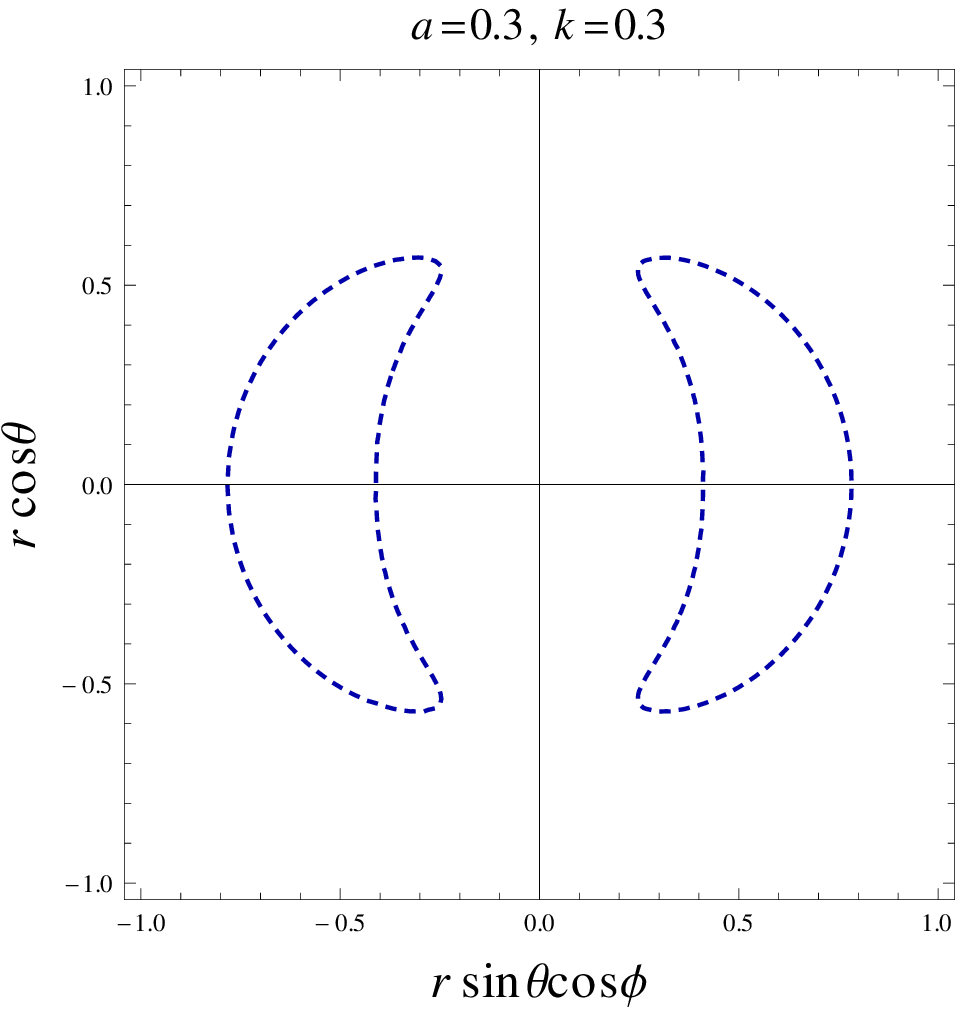}&
\includegraphics[scale=0.38]{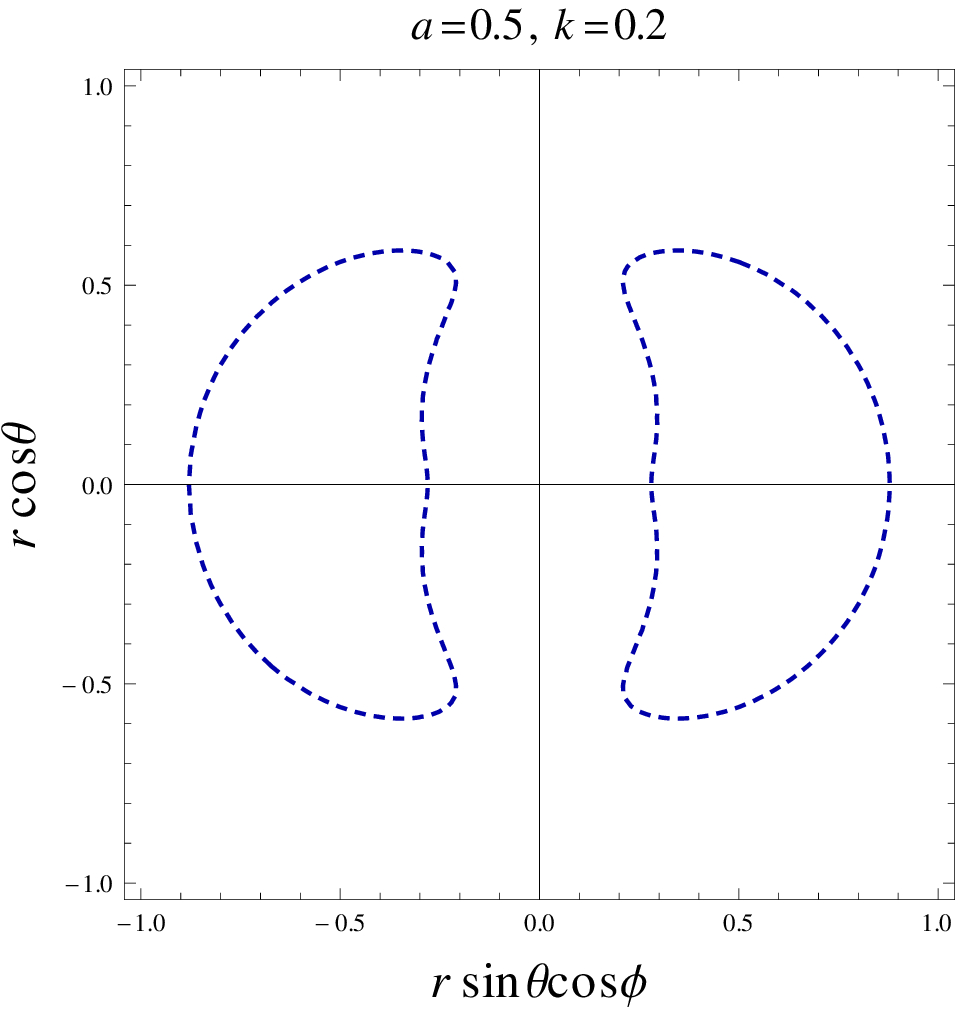}&
\includegraphics[scale=0.38]{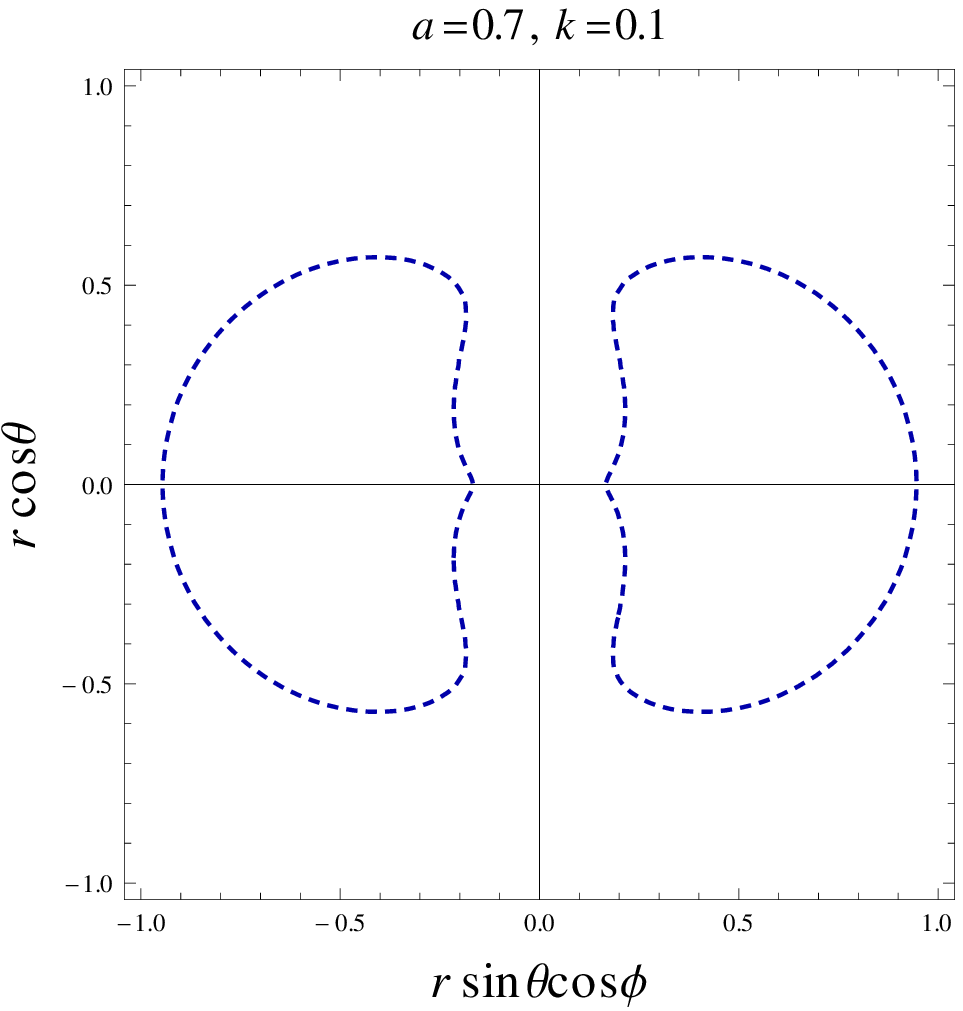}\\
\end{tabular}
\caption{Plots showing the behavior of ergoregion for the different values of spin parameter $a$ and deviation parameter $k$, the case $k=0$ corresponds to the $5D$ Kerr black hole.}\label{ergo}
\end{figure*}

\begin{table}
\caption{Table for values of the event horizon ($r_{+}$), Cauchey horizn ($r_{-}$) and $\delta = r_{+}-r_{-}$. \label{tb1} }
 \begin{center}
 \begin{tabular}{l l l l  l l l }
 \hline 
 &\multicolumn{3}{c}{$a=0.1$}  & \multicolumn{3}{c}{$a=0.3$} \\
 \hline  
$k$ & $r_{+}$ & $r_{-}$ & $\delta$ & $r_{+}$ & $r_{-}$ & $\delta$ \\
  \hline
0.0  \,\,  & 0.99498   &    \,\,\, -  & 0.99498      \,\, & 0.95393  & \,\,\,-  & 0.95393   \\
0.05  \,\, & 0.96855  & 0.11532   & 0.85322        \,\, & 0.92361  & 0.15141  & 0.77220 \\
0.15  \,\, & 0.90745  & 0.23414    & 0.67331        \,\, & 0.85001  & 0.29407  & 0.55594 \\
0.25  \,\, & 0.82690  & 0.35314     & 0.47276        \,\, & 0.73398  & 0.45222  & 0.28175  \\
$k_E$  \,\,& 0.60642 & 0.60642   & \,\,\,0        \,\, & 0.59935  & 0.59935  & \,\,\,0      \\
 \hline 
  \end{tabular}
 \end{center}
 \end{table} 
\begin{figure*}[h]
\begin{tabular}{c c}
\includegraphics[scale=0.45]{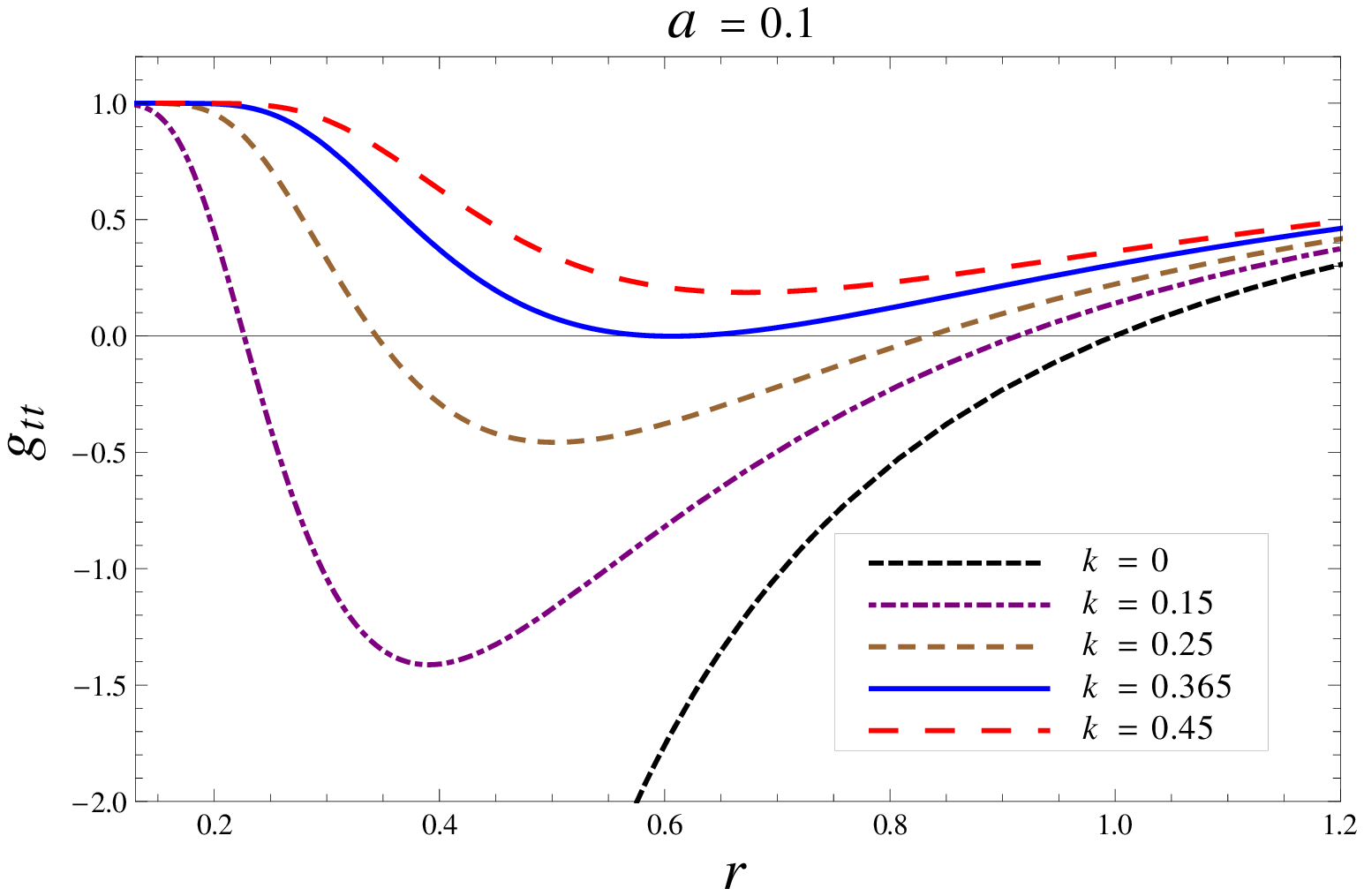}&
\includegraphics[scale=0.45]{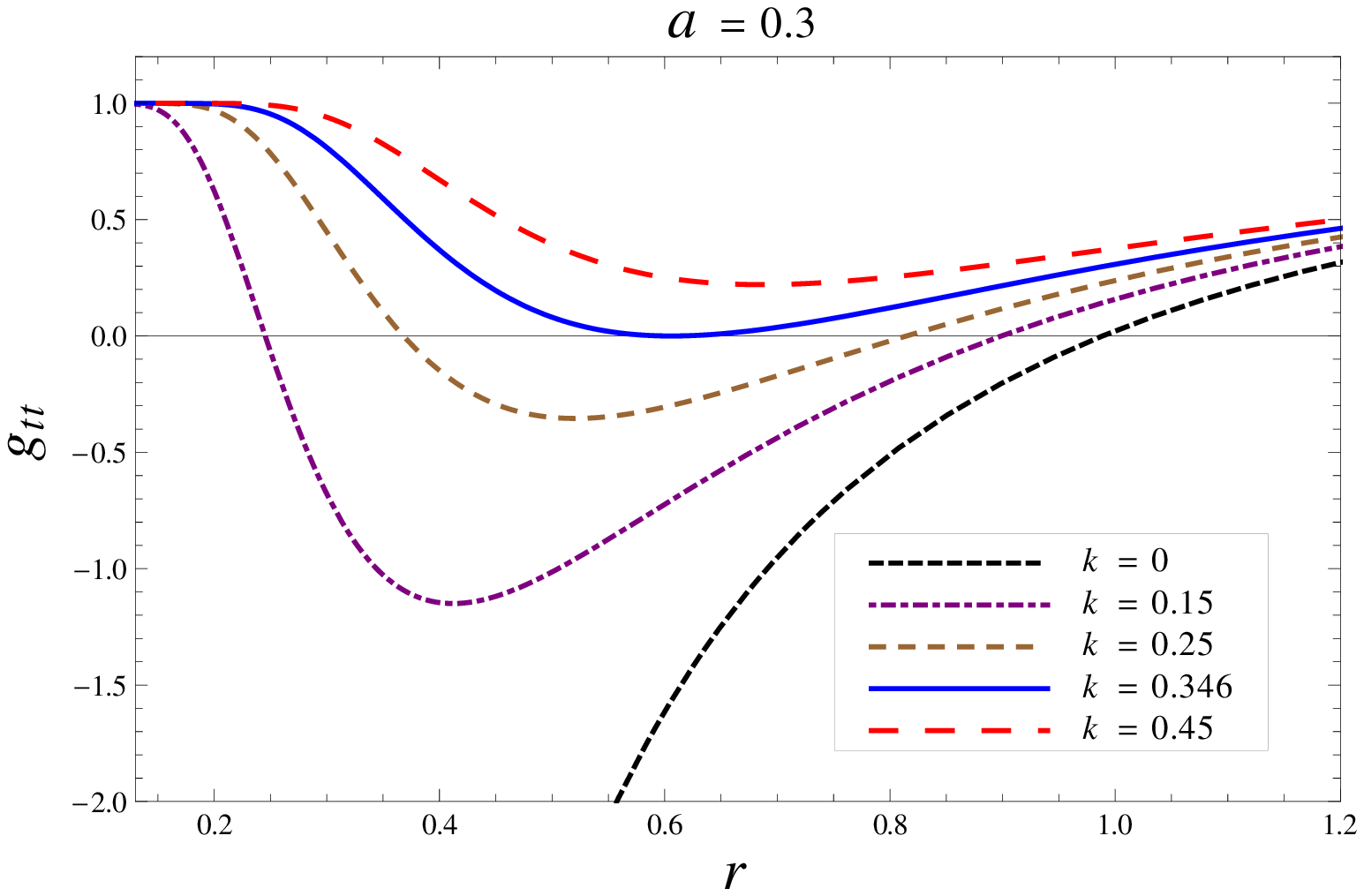}
\end{tabular}
\caption{Plots showing the behaviour of $g_{tt}$ vs radius $r$ for the
different values of deviation parameter $k$, the case $k=0$ corresponds to the $5D$ Kerr black hole.}
\label{slsf1}
\end{figure*}
The numerical analysis of the transcendental equation $\Delta = 0$ reveals that it is possible to find non-vanishing values of parameters $a$ and $k$ for which $\Delta$ has a minimum, and it admits two positive roots $r_{\pm}$ (see Table~\ref{tb1}). We find that, for a given $a$, there exists a critical value of $k$ is $ k_E$, and one of $r$ is $r_E$, such that $\Delta = 0$ admits a double root which corresponds to a regular extremal black hole with degenerate horizons $r_{-}= r_+ =r_E$. When $k < k_E$, $\Delta = 0$ has two simple zeros, and have no zeros for $ k > k_E$ (cf. Fig. \ref{ehf}). These two cases, respectively, to a $5D$ regular non-extremal black hole with a Cauchy horizon and an event horizon and a $5D$ regular spacetime. Indeed, the value of $k_E$ decreases with an increase in the value of rotation parameter $a$. \\
The static limit surface, where any observer cannot remain at rest and cannot be static or where the $g_{tt}$ component of the metric becomes zero. It requires the pre factor of $dt^2$ to be vanish 
\begin{eqnarray}\label{static}
g_{tt}=r^2- M e^{-k/r^2}+ a^2 \cos^2 \theta=0.
\end{eqnarray}
We have plotted the behaviour of static limit surface with radius $r$ in Fig.~\ref{slsf1}, which shows that for a given value of rotation parameter $a$ and angle $\theta$, disseminate a critical value of deviation parameter $k_s=0.365, 0.346$ for $a=0.1$ and $a=0.3$. Eq.~(\ref{static}) has no root if $k>k_s$, and have two simple zeros, if $k<k_s$ (cf. Fig.~\ref{slsf1}). Interestingly, the radii of an event horizon and static limit surface for the solution decrease when compared to the analogous Kerr case ($k = 0$). Note, for $\theta = 0$, the static limit surface and event horizon coincides as evident from Eq.~(\ref{static}). The static limit surface lies outside the event horizon and the region between the event horizon and static limit surface is known as an ergoregion. The ergoregion lies outside the black hole, it is possible to enter and leave again from the ergoregion. The behaviour of an ergoregion in $x-z$ plane is depicted in Fig.~\ref{ergo}, and shown that the area of the ergoregion depends on both $k$ as well as $a$. An increase in the value of $k$, result in the ergoregion increase or vice-versa. The numerical value of event horizon, static limit surface and ergoregion are listed in Table \ref{tb1} and \ref{tb2}.

\begin{table}
\caption{Radius of event horizons, static limit surfaces and $\delta^{a}=r_{SLS}-r_{+}$ for different values of parameter $k$.\label{tb2}}
 \begin{center}
 \begin{tabular}{l l l l  l l l l l l}
 \hline \hline
 &\multicolumn{3}{c}{$a=0.1$}  & \multicolumn{3}{c}{$a=0.3$} & \multicolumn{3}{c}{$a=0.5$}\\
 \hline 
$k$ & $r_{+}$ & $r_{SLS}$ & $\delta^{0.1}$ & $r_{+}$ & $r_{SLS}$ & $\delta^{0.3}$ & $r_{+}$ & $r_{SLS}$ & $\delta^{0.5}$   \\
  \hline
0.0  \,\, & 0.99498  & 0.99874  & 0.00376  \,\,& 0.95393  & 0.98868  & 0.03474  \,\,&  0.86602  & 0.96824 & 0.10222 \\
0.05  \,\, & 0.96855  & 0.97263  & 0.00408  \,\,& 0.92361  & 0.96170  & 0.03808  \,\,&  0.82402  & 0.93936 &  0.11534\\
0.15  \,\, & 0.90745  & 0.91249  & 0.00504  \,\,& 0.85001  & 0.89893  & 0.04891  \,\,&  0.69503  & 0.87060 &  0.17557\\
0.25  \,\, & 0.82690  & 0.83402  & 0.00712  \,\,& 0.73398  & 0.81464  & 0.08063  \,\,&  -  & - &  - \\
 \hline \hline
  \end{tabular}
 \end{center}
 \end{table}

From Table \ref{tb2}, we see that the ergoregion increase with increases in the values of deviation parameter $k$ and spin parameter $a$. So, the deviation parameter $k$, plays a major role in determining the horizon structure of the $5D$ rotating regular black hole.

The shape of the ergoregion, therefore, depends on the rotation parameter $a$, and deviation parameter $k$. The Penrose process \cite{Penrose:1971uk} relies on the presence of an ergoregion which for the solution (\ref{metric}) grows with the increase of deviation parameter $k$ as well as the rotation parameter $a$.

\begin{figure*}[h]
    \begin{tabular}{c c}
        \includegraphics[scale=0.45]{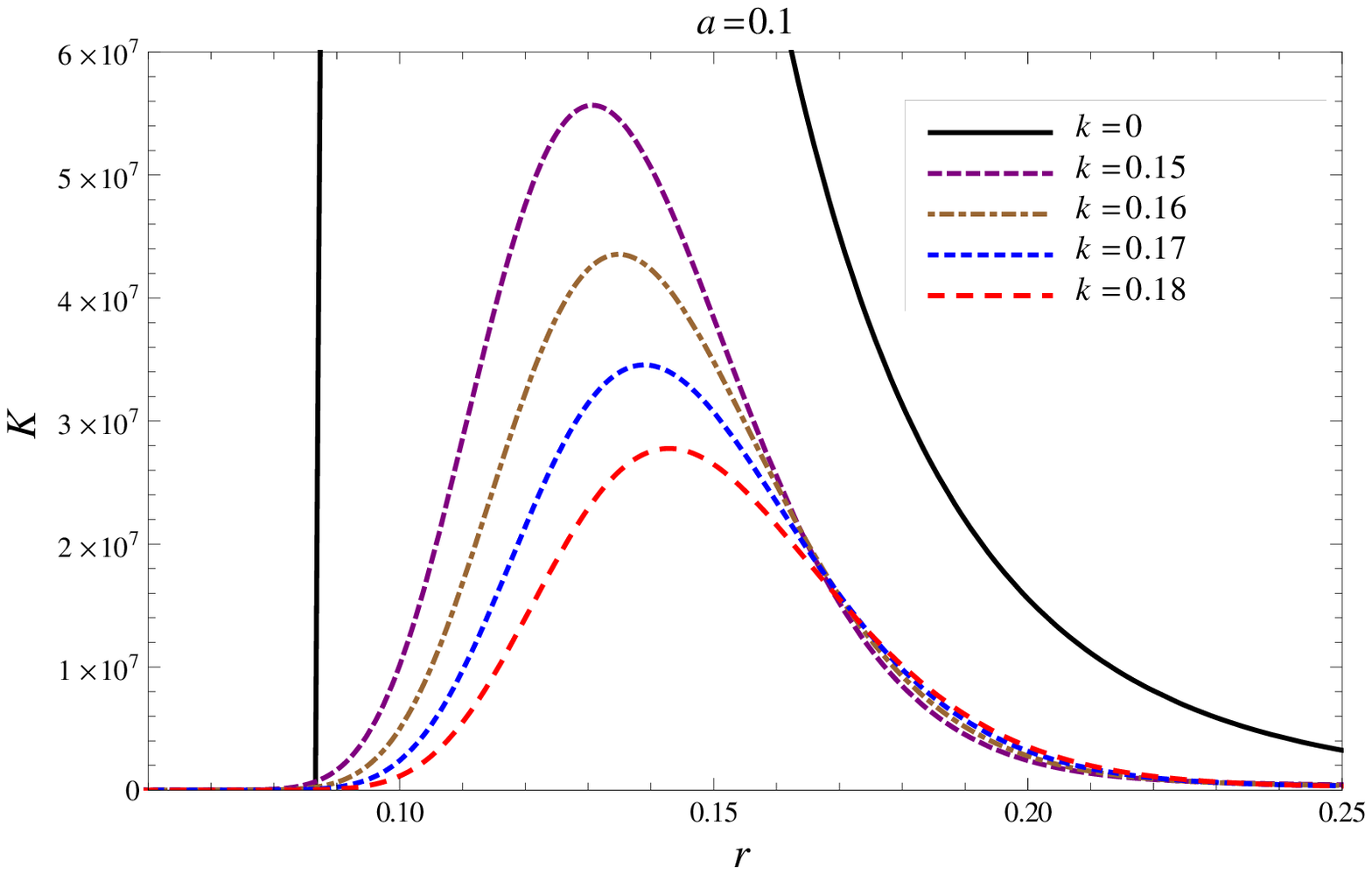} 
         \includegraphics[scale=0.45]{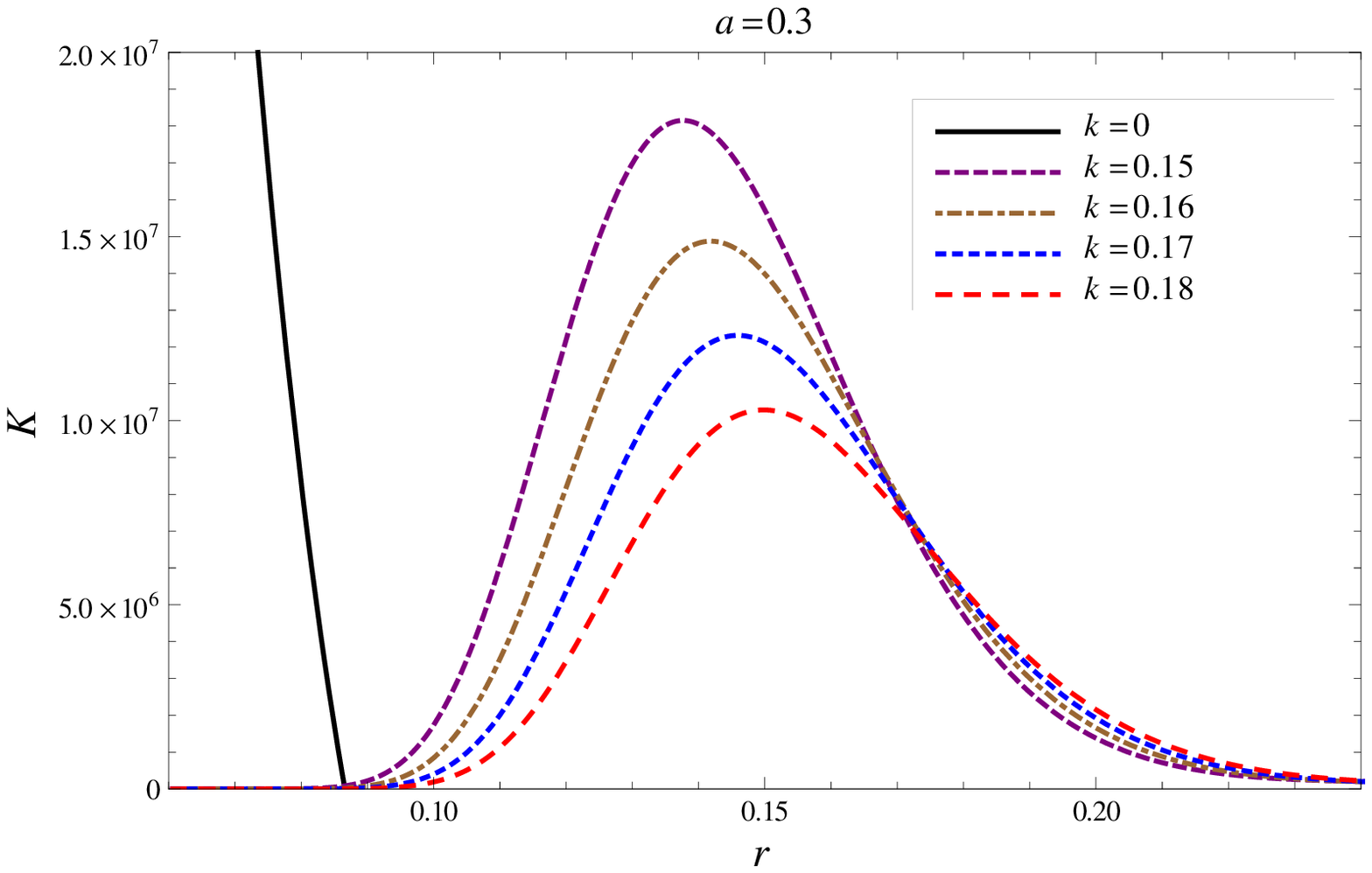}
    \end{tabular}
    \caption{\label{ks}Plots showing the regular behaviour of the Kretschmann scalar vs radius $r$ for the different values of deviation  parameters $k$.}
\end{figure*}

The regular behaviour of the metric (\ref{metric}), by plotting the Kretschmann scalar ($K=R_{abcd} R^{abcd}$) has shown in Fig.~\ref{ks}, which implies that the Kretschmann scalar is finite everywhere for $M \neq 0$ including $r=0$, so this solution is regular.  

The weak energy satisfies the relation $T_{ab}t^a t^b\geq0$ for all the timelike vectors $t^a$, i.e., energy density cannot be negative. The dominant energy condition requires $T^{ab} t^a t^b \geq 0$ and $T^{ab} t_b$ should be spacelike for any timelike vector $t^a$. Hence $\rho \geq0$ and $ \rho + P_i \geq 0$, are the requirement for satisfying the energy conditions, where $P_i$ are the momentum tensor. For $5D$ regular black hole, the weak energy conditions are written as
\begin{eqnarray}\label{wec}
&&\rho = \frac{3 M k e^{-k/r^2}}{\Sigma^3},\nonumber\\
&&\rho + P_2  = \frac{2 M k^2 e^{-k/r^2}}{r^2 \Sigma^3}= \rho + P_3 =\rho + P_4.
\end{eqnarray}
From Eq.~(\ref{wec}), it is clear that $5D$ rotating regular black hole satisfy the weak energy conditions.
\section{Particle Motion and black hole shadow}\label{pmbh}
An object casts a shadow, when it is situated between the observer and a light source. If a black hole is situated between a bright object like Quasars and observer, then it will cast a shadow. The apparent shape of a black hole is defined by the boundary of the shadow. To discuss black hole shadow, we have to study the motion of the particle. The equation of motion in the background of a $5D$ rotating regular black hole can be obtained by Lagrangian density
\begin{equation} \label{canonical}
\mathcal{L}= \frac{1}{2} g_{\mu \nu}u^{\mu}u^{\nu},
\end{equation}
where $u^{\mu}$ is the 5-velocity of the particle. The canonical momenta for the $5D$ rotating regular black hole are
\begin{eqnarray}\label{canonical}
p_{t}&=&\Big[-1+\frac{ M e^{-k/r^2}}{\Sigma}\Big]\dot{t}+\Big[\frac{ M a \sin^2 \theta e^{-k/r^2}}{\Sigma}\dot{\phi}\Big],\nonumber\\
p_{\phi}&=&\Big[\frac{ M a \sin^2 \theta e^{-k/r^2}}{\Sigma}\Big]\dot{t}+\Big[r^2+a^2+\frac{ M a^2 \sin^2 \theta}{\Sigma}\Big]\sin^2 \theta \dot{\phi},\nonumber\\
p_{\psi}&=&r^2 \cos^2 \theta \dot{\psi}, \quad
p_{r}=\frac{\Sigma}{\Delta}\dot{r},  \quad
p_{\theta}=\Sigma \dot{\theta}.
\end{eqnarray}
In Eq.~(\ref{canonical}), there are three conserved quantities corresponding to the energy $E$, and angular momentum $L_{\phi}$ and $L_{\psi}$ are given by $p_{t}=-E$,  $p_{\phi}=L_{\phi}$ and $p_{\psi}=L_{\psi}$. The equation of motion is obtained as
\begin{eqnarray}\label{tphi}
\dot{t}&=&\frac{1}{\Sigma \Delta}\Big[(\Delta \Sigma+  M (r^2+a^2) e^{-k/r^2}) E +  M a e^{-k/r^2} L_{\phi}\Big],\nonumber\\
\dot{\phi}&=&\frac{1}{\Sigma \Delta} \Big[- M a e^{-k/r^2} E + \frac{\Sigma- M e^{-k/r^2}}{\sin^2 \theta} L_{\phi}\Big], \nonumber\\
\dot{\psi}&=& \frac{L_{\psi}}{r^2 \cos^2 \theta}.
\end{eqnarray}
Next, the Hamilton-Jacobi equation is used to separate the radial and angular part of the equation of motion. The Hamilton-Jacobi equation for $5D$ black hole reads
\begin{eqnarray}\label{hamilton}
-\frac{\partial S}{\partial \lambda}=\frac{1}{2}g^{\mu \nu}\frac{\partial S}{\partial x^{\mu}}\frac{\partial S}{\partial x^{\nu}}
\end{eqnarray}
where $\lambda$ is an affine parameter, $g^{\mu \nu}$ is the metric tensor and $S$ is Jacobian action which takes the form
\begin{eqnarray}\label{action}
S=\frac{1}{2}m^2 \lambda - E t + L_{\phi} \phi + L_{\psi} \psi + S_{\theta} (\theta) + S_{r}(r),
\end{eqnarray}
where $m$ is the mass of the particle and $S_{r}(r)$, $S_{\theta}(\theta)$ are the function of $r$ and $\theta$.
Using the variable separable method and insert the Eq.~(\ref{action}) into the Eq.~(\ref{hamilton}), we obtain the equations of motion for photon ($m=0$) 
\begin{eqnarray}\label{stheta}
&&\Big(\frac{\partial S_{\theta}}{\partial \theta}\Big)^2 - E^2 a^2 \cos^2 \theta + \frac{L_{\phi}^2}{\sin^2 \theta}+ \frac{L_{\psi}^2}{\cos^2 \theta}-\mathcal{K}=0\nonumber\\
&&\Delta \Big(\frac{\partial S_r}{\partial r}\Big)^2 - E^2 r^2 -\frac{ M (r^2+a^2) e^{-k/r^2} E^2}{\Delta} - \frac{a^2 L_{\phi}^2}{\Delta}-\frac{2 a M e^{-k/r^2} E L_{\phi}}{\Delta} + \frac{a^2 L_{\psi}^2}{r^2}+\mathcal{K}=0,\nonumber\\
\end{eqnarray}
where constants $E$, $L_{\phi}$ and  $L_{\psi}$ are energy and angular momentum corresponding to the $\phi$ and $\psi$ axes, respectively, and  $\mathcal{K}$ is a constant of separation called Carter constant.  The Eq.~(\ref{stheta}) can be recast into the form
\begin{eqnarray}
\frac{\partial S_{\theta}}{\partial \theta} = \sqrt{\Theta(\theta)} \quad \mbox{and} \quad \frac{\partial S_r}{\partial r} = \sqrt{{\cal{R}}(r)},
\end{eqnarray}
where the expression $\Theta(\theta)$ and ${\cal{R}}(r)$ are given by
\begin{eqnarray}\label{theta}
&&\Theta(\theta) = E^2 a^2 \cos^2 \theta - \frac{L_{\phi}^2}{\sin^2 \theta} - \frac{L_{\psi}^2}{\cos^2 \theta} + \mathcal{K},\\
&&{\cal{R}}(r) = \Delta (E^2 r^2 - \frac{a^2 L_{\psi}^2}{r^2} -\mathcal{K})+  M (r^2+a^2)e^{-k/r^2} E^2 + a^2 L_{\phi}^2 + 2 a M e^{-k/r^2} E L_{\phi}.
\end{eqnarray}
These equations define the propagation of photon around the spacetime of the $5D$ rotating regular black hole. For a particle moving in the equatorial plane and to remain in the equatorial plane iff $\cal{K}$ = 0. One can recover the equation of motion for $5D$ Kerr-like black hole in the absence of a deviation parameter ($k = 0$). For obtaining the boundary of the black hole shadow it demands the study of radial equation. We can rewrite the radial equations of motion \cite{Papnoi:2014aaa}
\begin{equation}
\frac{1}{2} \dot{r}^2 + V_{eff} = 0,
\end{equation}
where 
\begin{eqnarray}\label{eff}
V_{eff} = -\frac{1}{2 \Sigma^2}\Big[\Delta(E^2 r^2 - \frac{a^2 L_{\psi}^2}{r^2} -\mathcal{K})+ M (r^2+a^2) e^{-k/r^2}E^2+ a^2 L_{\phi}^2 + 2 a M e^{-k/r^2} E L_{\phi}\Big] \nonumber\\
\end{eqnarray}
Let us define the following impact parameters $\xi_1=L_{\phi}/E$, $\xi_2=L_{\psi}/E$ and $\eta=\mathcal{K}/E^2$. Now we can rewrite the ${\cal R}(r)$ in terms of new parameters,
\begin{equation}\label{R1}
{\cal R}(r) = \Delta (r^2 - \frac{a^2 \xi_2^2}{r^2}-\eta)+ M (r^2+a^2)e^{-k/r^2} + a^2 \xi_1^2 + 2 a M e^{-k/r^2} \xi_1
\end{equation}
Here we are interested in the radial motion of the photons, which are essential for determining the shape of the shadow of a $5D$ rotating regular black hole. 
 The incoming photons, which are coming towards the black hole from a light source, when they reach near the black hole, then they follow the three possible trajectories, either they fall into the black hole or scattered away from the black hole or make a circular orbit near the black hole. The unstable circular orbit near the black hole is responsible for determining the shape of the shadow. We obtain the unstable circular photon orbits by the following conditions 
\begin{eqnarray}\label{circular}
&&V_{eff}=0 \quad \mbox{and} \quad \frac{d V_{eff}}{d r}=0, \nonumber\\
&& {\cal R}(r)=0 \quad \mbox{and} \quad \frac{d{\cal R}(r)}{d r}=0.
\end{eqnarray}
The impact parameters $\xi_1$ and $\eta$ determine the contour of the shadow for the photon orbits around the black hole. We solve the Eqs.~(\ref{R1}) and (\ref{circular}), and obtain the value of $\eta$ and $\xi_1$ for the $5D$ rotating regular black hole
\begin{eqnarray}\label{eta}
&&\eta=\frac{2 r^{10}+2 k  r^2 (k-2 r^2)e^{-2k/r^2}+a^2( k  e^{-k/r^2}+r^4)^2}{(- k  e^{-k/r^2}+r^4)^2},\nonumber\\
&&\xi_1=\frac{(a^2+r^2)(- k e^{-k/r^2}+r^4)+ 2  e^{-k/r^2}((a^2+r^2)k-r^4)}{a(- k e^{-k/r^2}+r^4)}
\end{eqnarray}
The case $k=0$ correspond to the $5D$ Kerr black holes, 
\begin{eqnarray}
&& \eta = 2 r^2 +a^2, \nonumber\\
&& \xi_1=\frac{r^2+a^2-2}{a}.
\end{eqnarray}

\noindent  To obtain the apparent shape of the black hole, we introduce the celestial coordinates $\alpha$ and $\beta$ \cite{Johannsen:2015qca},
\begin{eqnarray}\label{alpha}
\alpha &=& \lim_{r_0 \to \infty} - r_0 \frac{(p^{\phi}+p^{\psi})}{p^{t}}, \nonumber\\
\beta &=& \lim_{r_0 \to \infty} r_0 \frac{p^{\theta}}{p^t}.
\end{eqnarray}
We take the limit $r \rightarrow \infty$, because the distance between the observer and the black hole is large and $\theta_0$ is the angular coordinate of the observer. For $5D$ rotating regular black hole, celestial coordinates take the form \cite{Amir:2017slq}
\begin{eqnarray}\label{beta}
\alpha &=& -\Big(\xi_1   \csc \theta_0 +\xi_2 \sec \theta_0 \Big), \nonumber\\
\beta &=& \pm \sqrt{\eta-\xi_1^2 \csc^2 \theta_0 - \xi_2^2 \sec^2 \theta_0 + a^2}.
\end{eqnarray}
 We choose an equatorial plane for observing the shadow of the black hole, so the angle of the inclination is $\theta_0 = \pi/2$. From Eq.~(\ref{canonical}), if $\theta =\pi/2$, then $L_{\psi}=0$, which implies $\xi_2=0$, so the Eq.~(\ref{beta}) reduces into the form
\begin{eqnarray}\label{beta1}
\alpha &=& - \xi_1, \nonumber\\
\beta &=& \pm \sqrt{\eta - \xi_1^2 + a^2}.
\end{eqnarray}
The shadows of a $5D$ rotating regular black hole can be visualized by the plotting of $\alpha$ vs $\beta$ for the different values of the rotation parameter ($a$) and deviation parameter ($k$) at different inclination angles. One can check the relation of celestial coordinates $\alpha $ and $\beta$  in Eq.~(\ref{beta1}) and find that they follow the relation $\alpha^2+\beta^2=\eta+ a^2$, where $\eta$ is given by Eq.~(\ref{eta}). Eq.~(\ref{beta1}) depends upon the values of deviation parameter $k$ and rotation parameter $a$ and also due to extra dimension. The shadow for a $5D$ non-rotating regular black hole can be obtained from
\begin{equation}
\alpha^2+\beta^2= \frac{4}{2-e^{-k/2}}.
\end{equation}
\begin{figure*}[ht]
\includegraphics[scale=0.42]{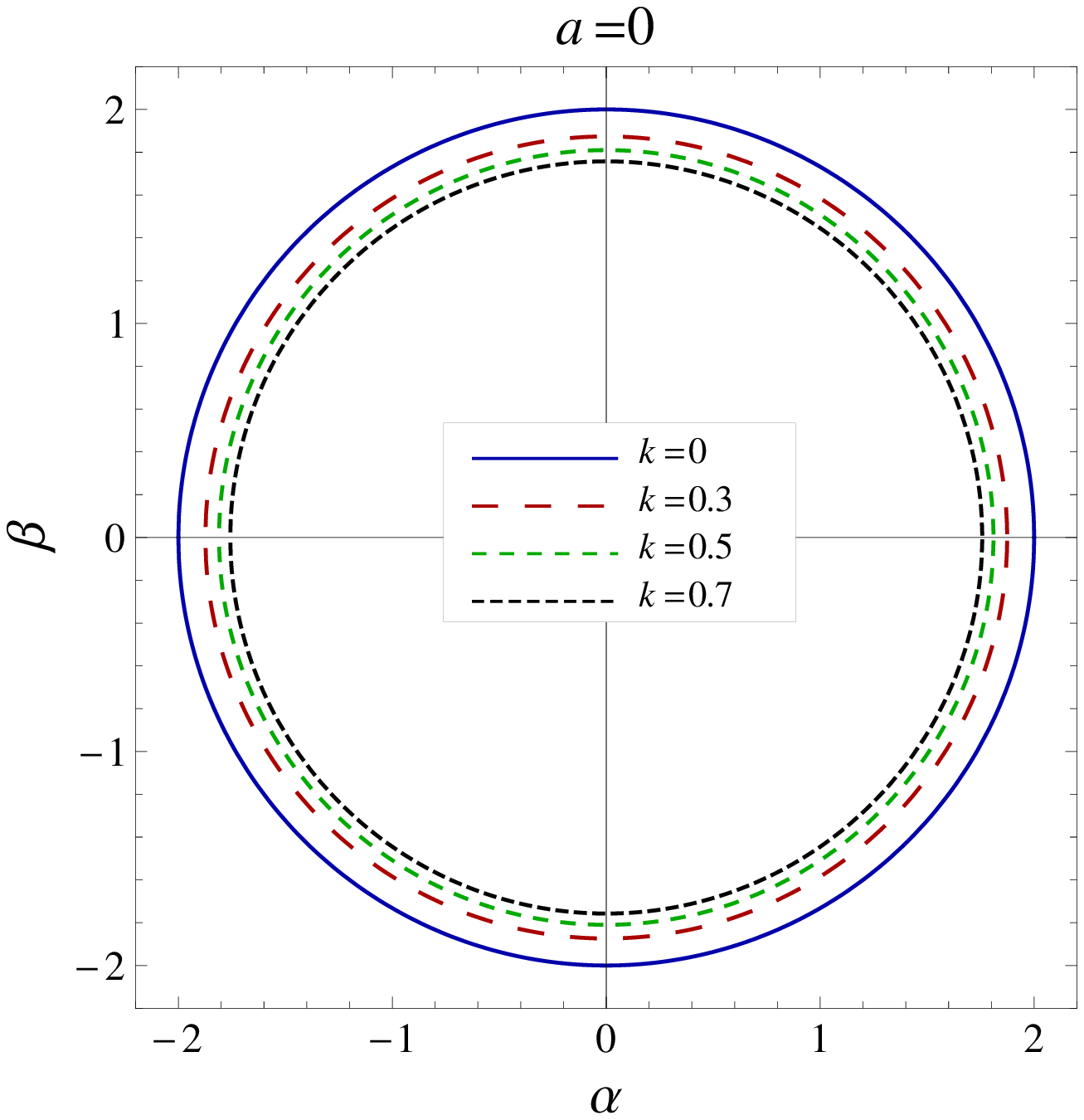}
\includegraphics[scale=0.55]{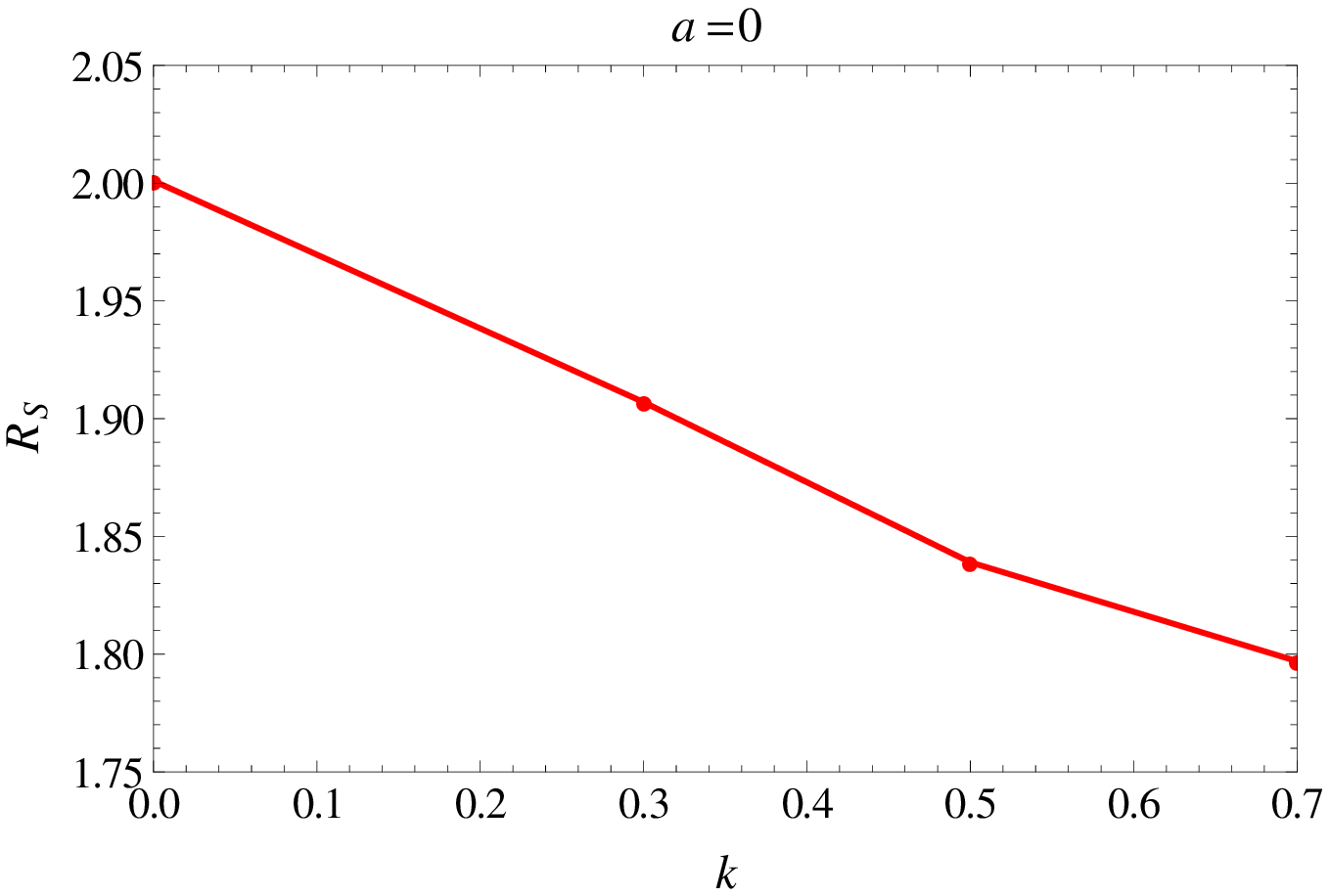}
\caption{Plot showing the shapes of the black shadow for non-rotating $5D$ regular black hole with different values of deviation parameter $k$.}
\label{shad1}
\end{figure*}

The $5D$ non-rotating regular black hole is a general case of $5D$ Schwarzschild-Tangherlini  black hole and its shadow appears as a perfect circle with radius $R_s$ depicted in Fig.~\ref{shad1}. We plotted the shadow of a $5D$ non-rotating regular black hole for several values of deviation parameter $k$. The effect of deviation parameter has shown with the radius of the circle in Fig.~\ref{shad1}. Thus the size of the shadow decreases with increasing the value of $k$.

Now we investigate the shape of the shadow for the $5D$ rotating regular black hole. The celestial coordinates in the rotating case also satisfy the relation, $\alpha^2+\beta^2=\eta+ a^2$. If we calculate the celestial coordinate relation in the absence of charge, i.e., $k = 0$
\begin{equation}
\alpha^2+\beta^2=2(r^2+a^2)
\end{equation}
\begin{figure*}[h]
\includegraphics[scale=0.45]{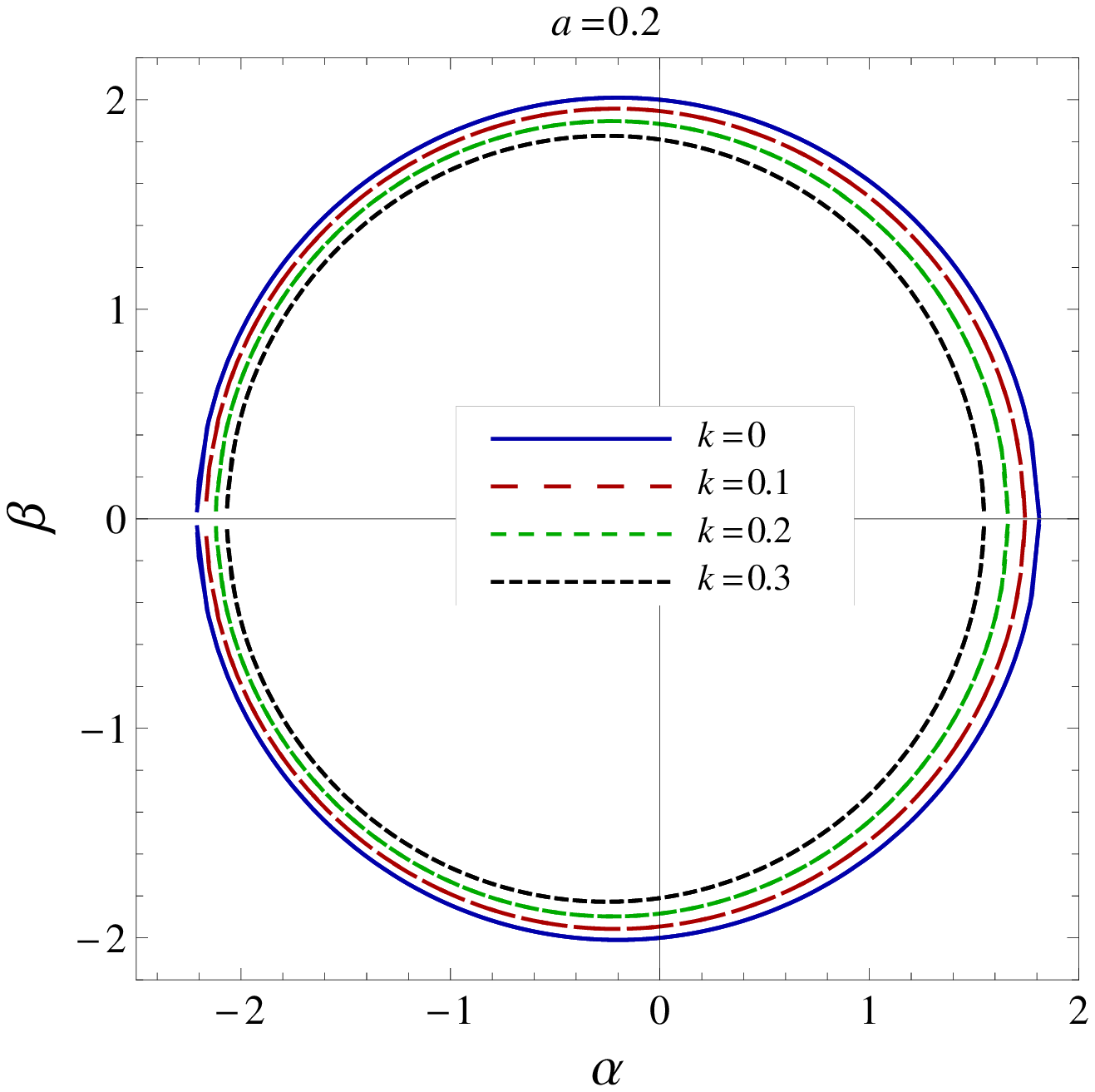}
\includegraphics[scale=0.45]{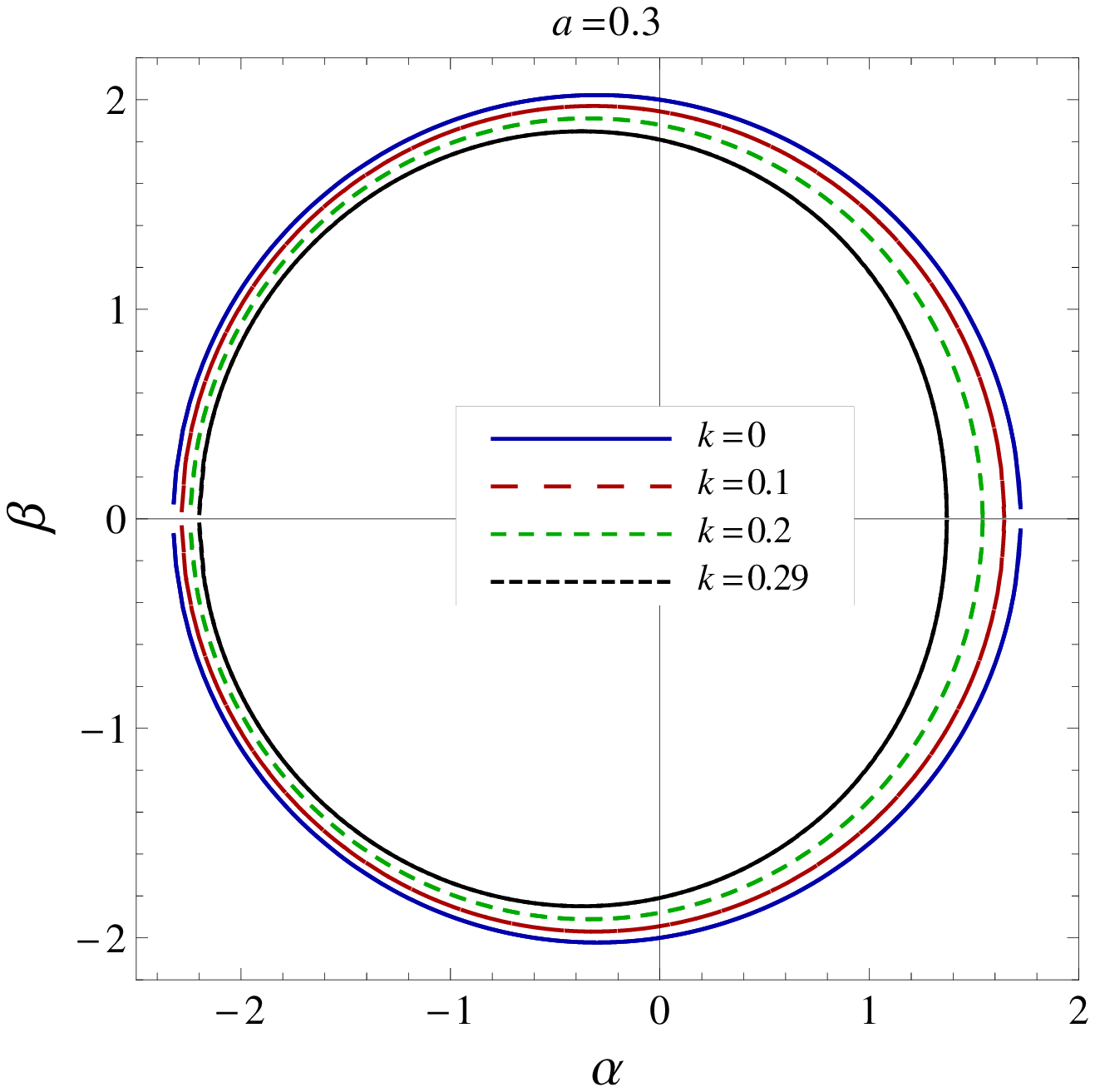}
\includegraphics[scale=0.45]{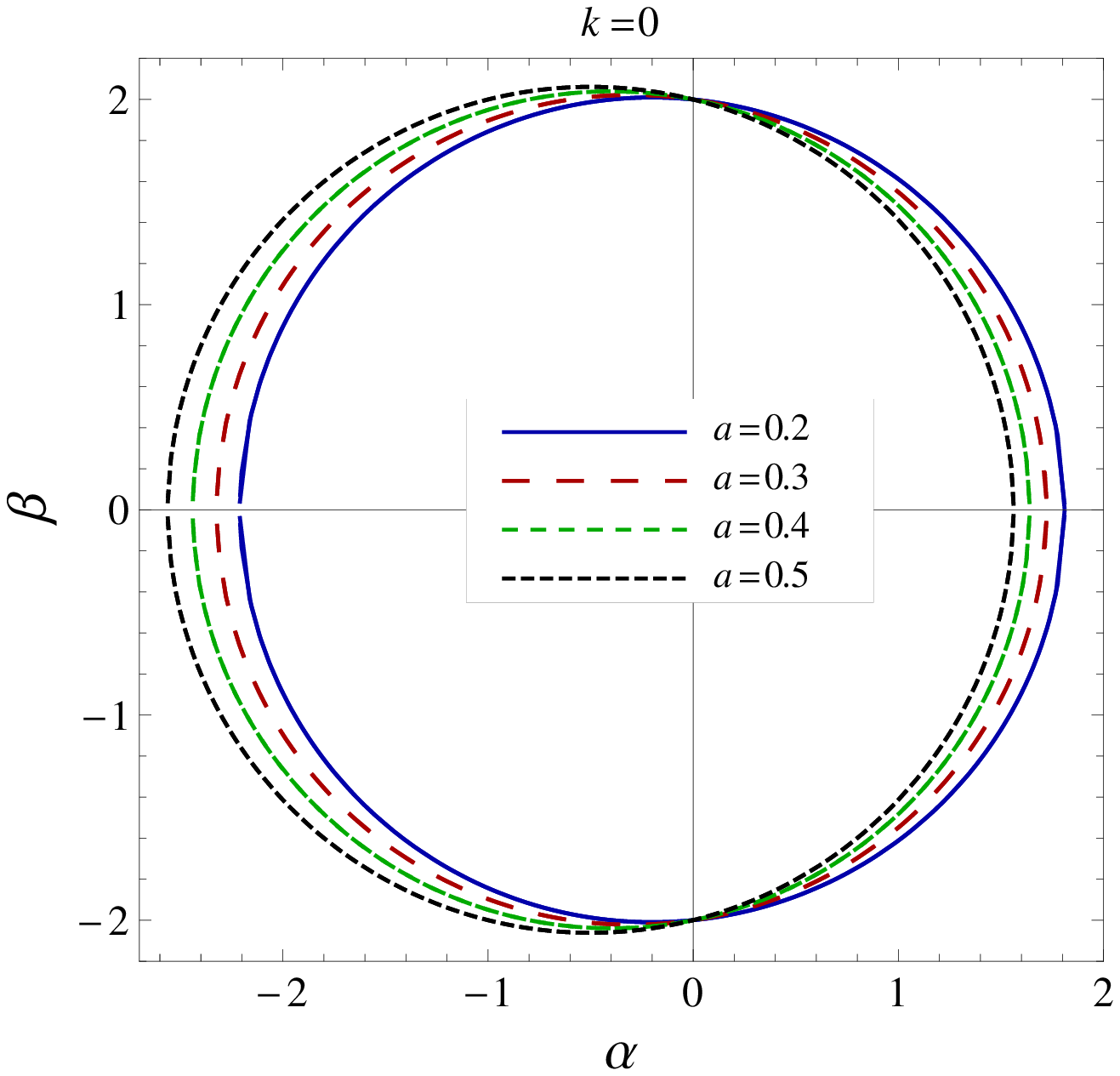}
\includegraphics[scale=0.45]{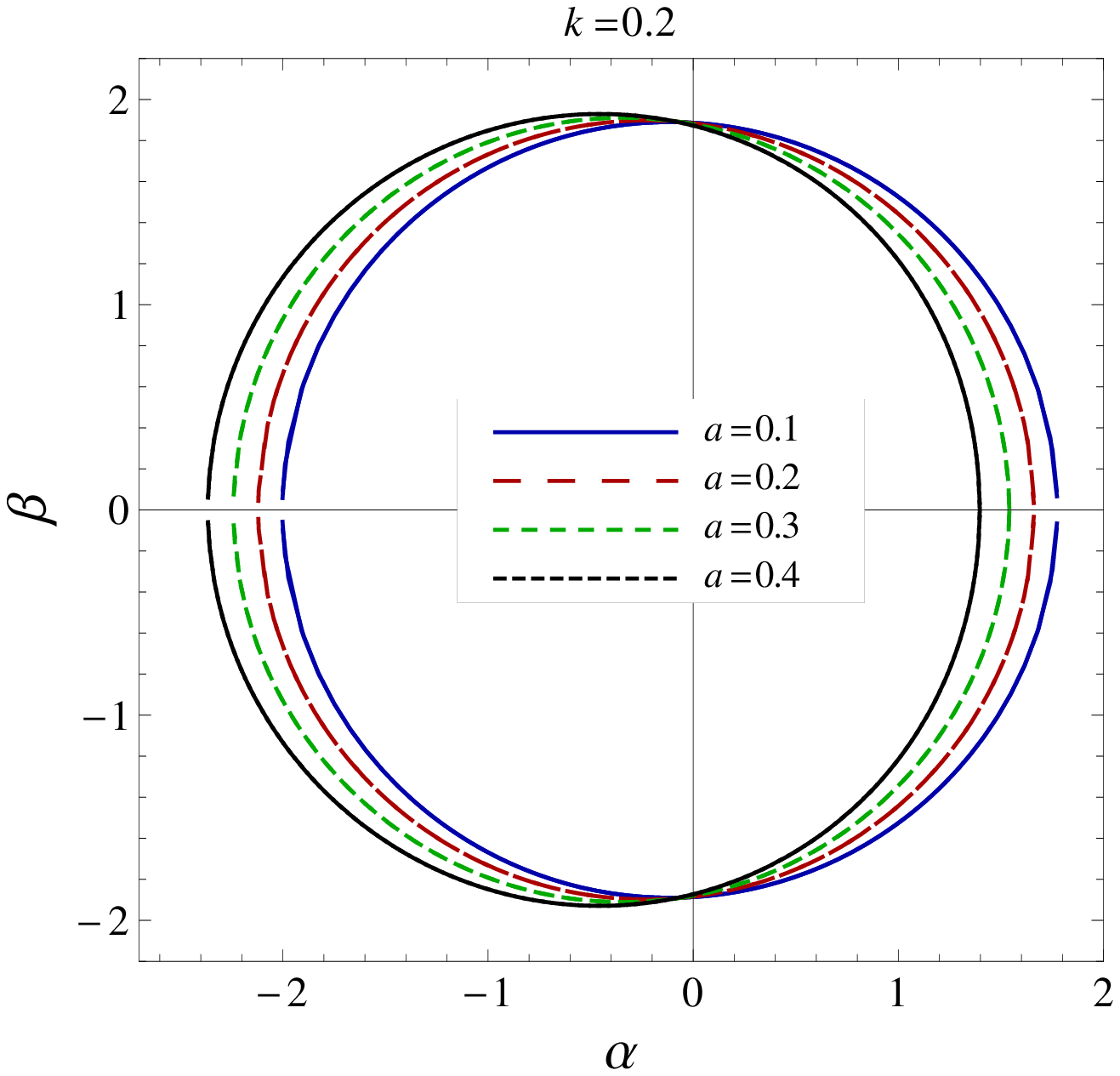}
\caption{Plots showing the shapes of the black hole shadow for rotating $5D$ regular black hole with different values of rotation parameter $a$ and the deviation parameter $k$.}
\label{shad2}
\end{figure*}
The shape of the shadow of the $5D$ rotating regular black hole have been plotted in Fig.~\ref{shad2} for the different values of deviation parameter $k$ and rotation parameter $a$. The shape of the black hole shadow is not a perfect circle. We observe that the shape of the shadow has affected due the parameters $a$, $k$ and extra dimension. The size of the shadow decreases continuously (cf. Fig.~\ref{shad2}) when we increase the value of $k$. This can be understood due to the dragging effect of a rotating black hole.  

\begin{figure}
	\includegraphics[scale=0.5]{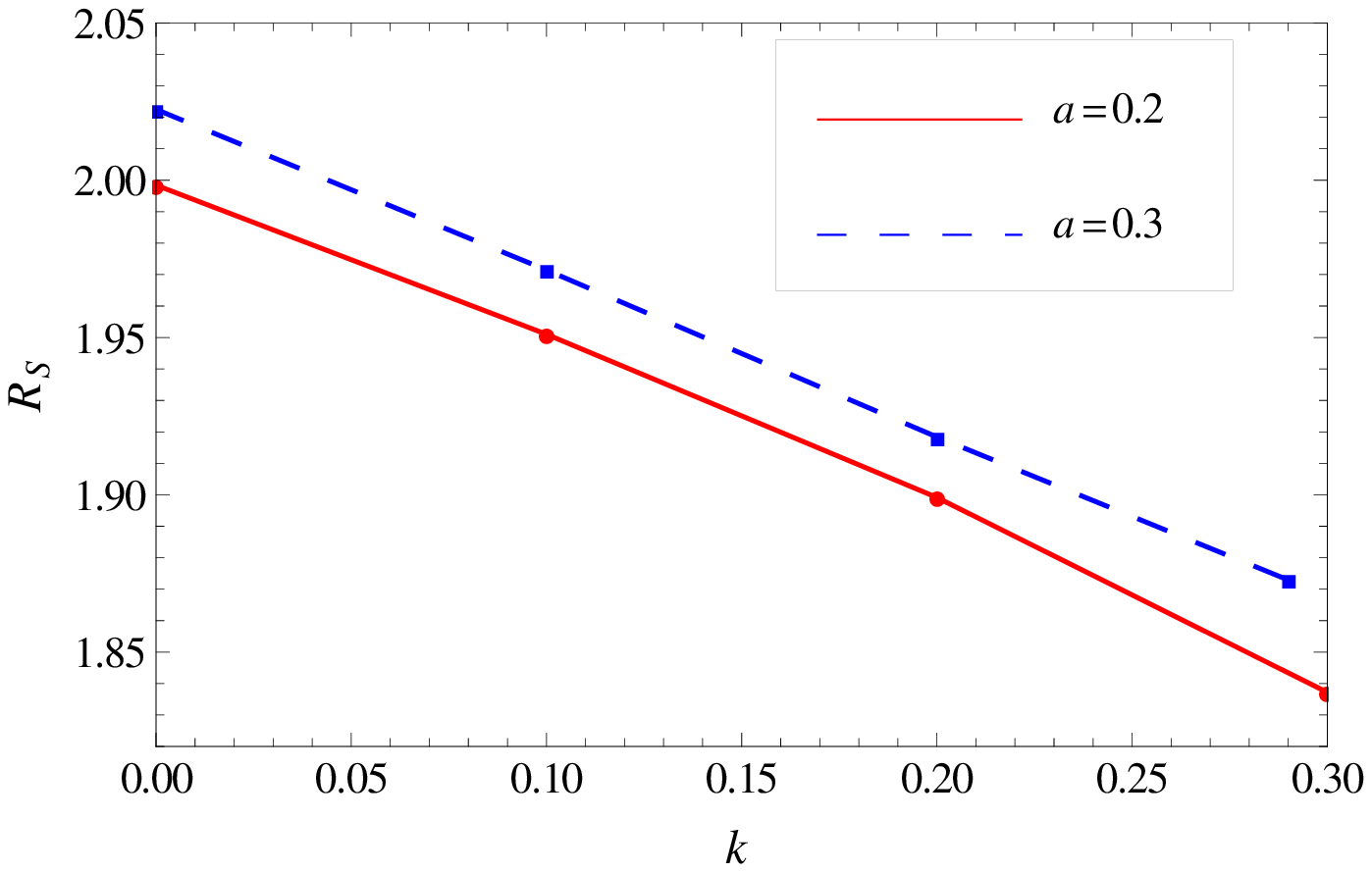}
	\includegraphics[scale=0.5]{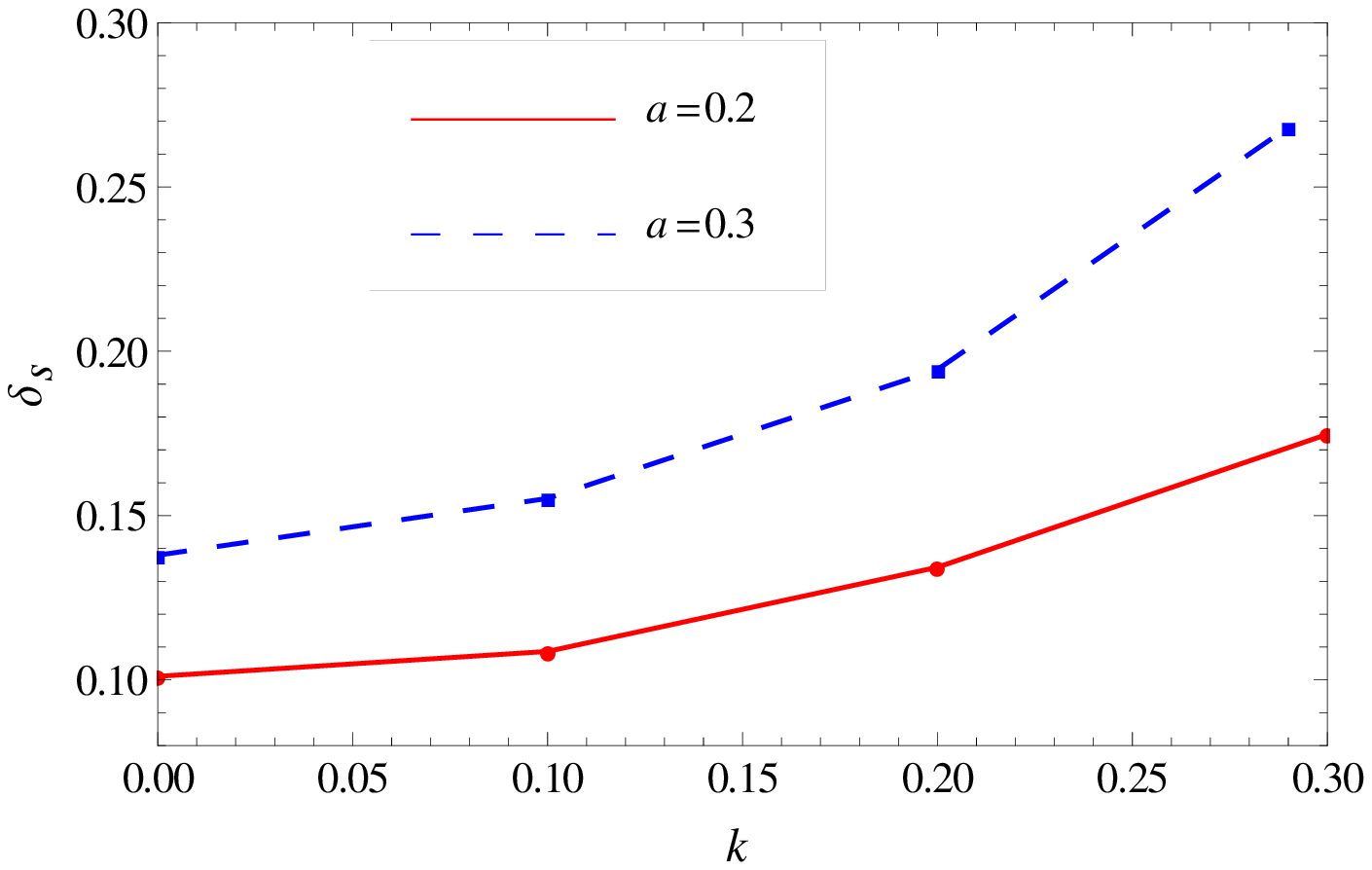}
    \caption{ Plots showing the variation of radius of shadow $R_s$ and distortion parameter $\delta_s$ with deviation parameter $k$ for different values of spin $a$.}
\label{obs}
\end{figure}

Next, we introduce some other observables,  which are helpful, when we study the radius and the distortion of the shadow. We consider the Hioki-Maeda characterization \cite{Hioki:2009na} to determine the observable, i.e., radius $R_s$ and deformation $\delta_s$, for $5D$ rotating regular black hole
\begin{eqnarray}
R_s &=& \frac{(\alpha_t - \alpha_r)^2 + \beta_t^2}{2(\alpha_t - \alpha_r)}, \\
\delta_{s} &=& \frac{\tilde{\alpha_p} -\alpha_p}{R_s},
\end{eqnarray}
\begin{figure}[h]
	\includegraphics[scale=0.3]{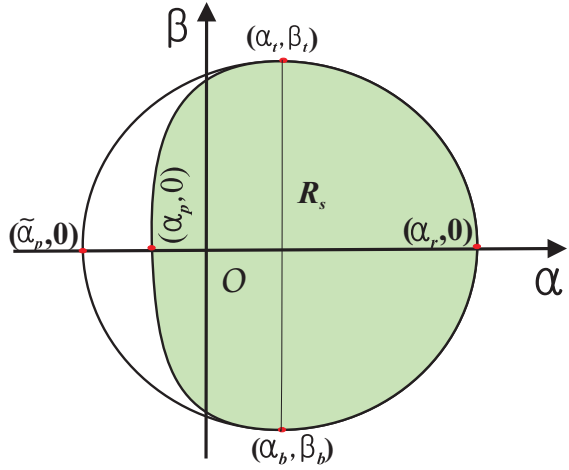}
    \caption{\label{observable} Schematic representation of the observables for rotating black holes
    \cite{Amir:2016cen}.}
\end{figure}
where $(\tilde{\alpha_p} ,0)$ and $(\alpha_p,0)$ are the coordinates of the given reference circle 
and the contour of the shadow, which cut the horizontal axis of the black hole from the opposite side of $(\alpha_r,0)$ (cf. Fig.~\ref{observable}). In this method, we choose a reference circle which 
passes through the three coordinates of the shadow of a black hole. We choose the positions of the coordinates at the top ($\alpha_t$, $\beta_t$), bottom ($\alpha_b$, $\beta_b$), and rightmost position ($\alpha_r$, 0) 
(cf. Fig.~\ref{observable}). We plotted the behaviour of the radius $R_s$ and distortion $\delta_s$ with deviation parameter $k$
for different values of $a$ in Fig.~\ref{obs} which suggested that 
the radius $R_s$ of the shadow and the distortion $\delta_s$ depend upon the values of $k$ as well as $a$.

\subsection{Energy Emisson Rate}\label{eert}
In this section, we will discuss the rate of energy emission from the $5D$ rotating regular black hole (\ref{metric}). The low energy absorption cross-section for a spherically symmetric black hole is always the area of the horizon \cite{41}. However, the absorption cross section oscillates around a limiting value at high energy scale, which took the value of geometrical cross section $\sigma_{lim}$ of the photon sphere around the black hole \cite{Wheeler} and takes the value
\begin{equation}
\sigma_{lim}=\pi R_s^3,
\end{equation}
where $R_s$ is black hole shadow radius. The expression of the energy emission rate of black hole reads \cite{Mashhoon:1973zz,Wheeler}
\begin{equation}
\frac{d^2E{\nu}}{d\nu dt}= \frac{4 \pi^3 R_s^3 }{e^{\omega/T_{+}}-1}\omega^3,
\end{equation}
where $\omega$ is the frequency of the rotation and $T_{+}$ is the Hawking temperature at the outer horizon and given by 
\begin{eqnarray}
T_{+}=\frac{1}{2 \pi (r_{+}^2+a^2)}\left[r_{+}-\frac{k(r_{+}^2+a^2)}{r_{+}^3} \right].
\end{eqnarray}

\begin{figure}
	\includegraphics[scale=0.53]{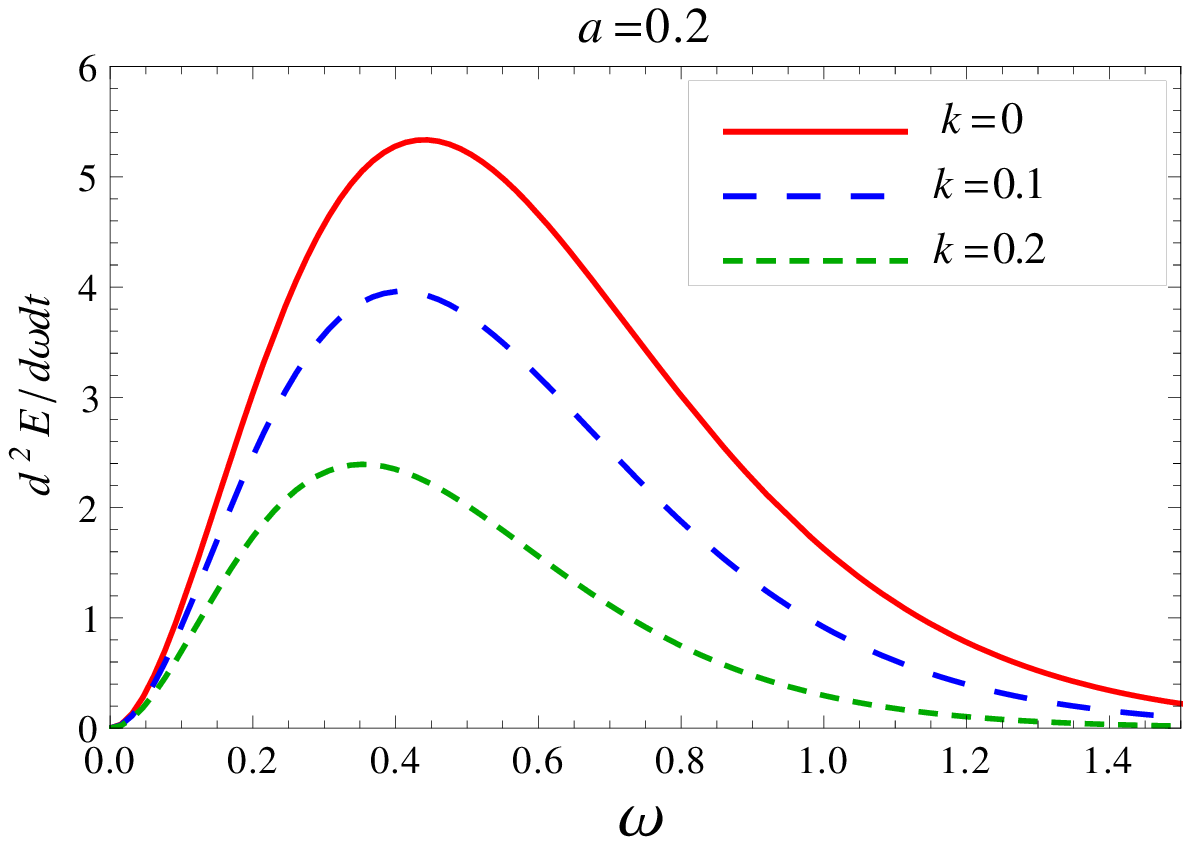}
	\includegraphics[scale=0.53]{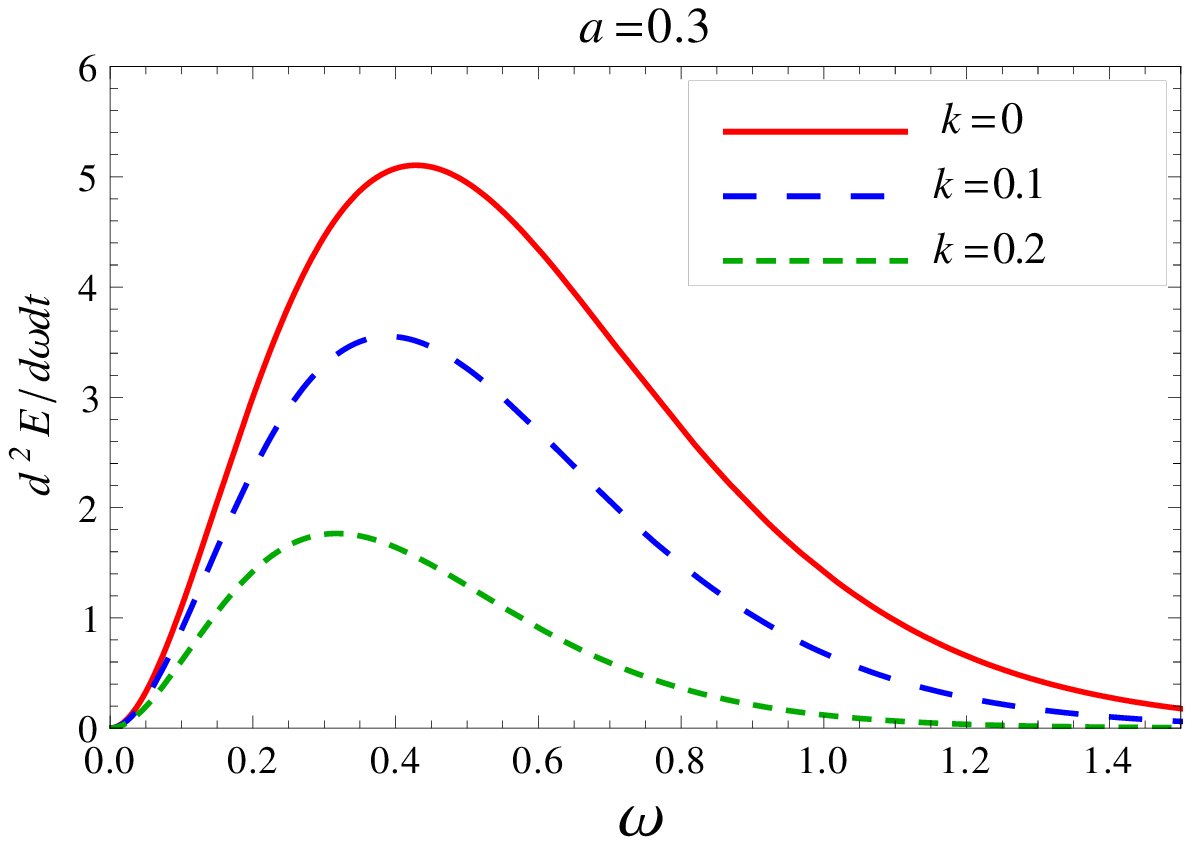}
    \caption{ Plots showing the variation of the energy emission rate with frequency $\omega$ for different values of $k$ and $a$.}
\label{et11}
\end{figure}

 We plot the energy emission rate with frequency $\omega$ for the different values of $a$ and $k$ in Fig.~\ref{et11}.

 \section{Conclusion}\label{cnbh}
We have presented rotating regular black holes to the $5D$  Einstein gravity coupled to nonlinear electrodynamics. We have tried to show that a simple idea  may provide  $5D$ regular black holes which go over to $5D$ Kerr black holes when $k=0$ and behaves asymptotically $r \gg k$ like  $5D$ Kerr-Newman black holes. The obtained metric depends on an additional parameter $k\geq 0$ which is related to mass $M$ and charge $e$ of the black hole  and depending on its value one gets black holes with two horizons or extremal black holes.  We have studied and plotted the horizons and ergoregions to explicitly bring out the effect of parameter $k$.

 We have also studied the shadow casted by the obtained solution and obtained exact expressions for observables. The solution allows us to distinguish the shadow associated with the $5D$ Kerr black holes that one of the $5D$ rotating regular black holes. It turns out that the parameter $k$ affects the horizons, ergoregion, and also black hole shadow. Indeed, with the increase in the parameter $k$, the size of the shadow decreases and shape gets more deformed.

In view of recent observations of event horizon telescope, studies of the $5D$ regular black hole is an important and happen to be timely.  Our analysis suggests very rich spacetime properties due to the parameter $k$. The sudy of a $5D$ regular black hole shadow may be helpful at better understanding of observation from event horizon telescope.

\section{Acknowledgement}
S.G.G. would like to thanks SERB-DST Research Project Grant No. SB/S2/HEP-008/2014 and DST INDO-SA bilateral project DST/INT/South Africa/P-06/2016.
\appendix
\section {Thermodynamics of $5D$ Regular Black Hole}

\begin{figure*}[h]
\begin{tabular}{c c}
\includegraphics[scale=0.6]{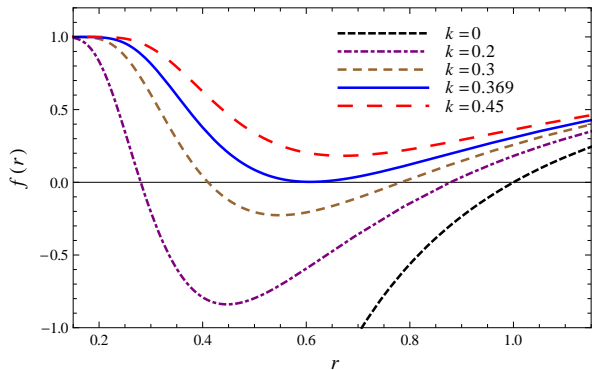}
\end{tabular}
\caption{Plots showing the behaviour of metric function $f(r)$ vs radius $r$ for the different values of deviation parameter $k$.}\label{slsf4}
\end{figure*}

The solution (\ref{eqn.gb}) can also be understood as a black hole of general relativity coupled to nonlinear electrodynamics.  Now, we look for the event horizon, which we obtain by $g^{rr}=1-{2Me^{-k/r^2}}/{r^2}=0$. The numerical analysis of $f(r_E)=0$ shows that it admits two roots $r_{\pm}$. The smaller and larger roots, respectively, corresponds to the Cauchy and event horizons. The radius of the event horizon decreases with an increase in the value of deviation parameter $k$ as shown in Fig.~\ref{slsf4}.

Next, we study the thermodynamic quantities associated with the $5D$ regular black hole solution (\ref{eqn.gb}) in terms of the horizon radius $r_+$. The mass of the black hole can be obtained by using using (\ref{eqn.gb})
\begin{equation}
M_+=r_+^2e^{k/r_+^2}
\label{eqM}
\end{equation}
When we put  $k=0$ in Eq.~(\ref{eqM}), then it reduces to the mass of the $5D$ Schwarzschild-Tangherlini black hole. The Hawking temperature of the black hole is proportional to the surface gravity  $\kappa$ by $T=\kappa/2\pi$, where $\kappa$  is $ \kappa=\frac{1}{2\pi}\left(-{1}\nabla_{\mu}\xi_{\nu}\nabla^{\mu}\xi^{\nu}/2 \right)^{1/2}$, where $\xi^{\mu}=\partial/\partial t$ is a Killing vector. The temperature of the $5D$ regular black hole is
\begin{equation}
T_+=\frac{1}{2\pi r_+}\left[1-\frac{k}{r_+^2}\right].
\label{eqT}
\end{equation}

\begin{figure*}[h]
\begin{tabular}{c c}
\includegraphics[scale=0.6]{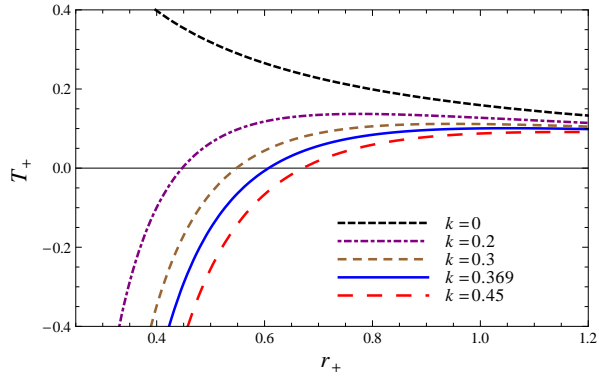}
\end{tabular}
\caption{Plots showing the behaviour of temperature $T_+$ vs horizon radius $r_+$ for the different values of deviation parameter $k$.}\label{slsf5}
\end{figure*}
\begin{figure*}[h]
\begin{tabular}{c c}
\includegraphics[scale=0.6]{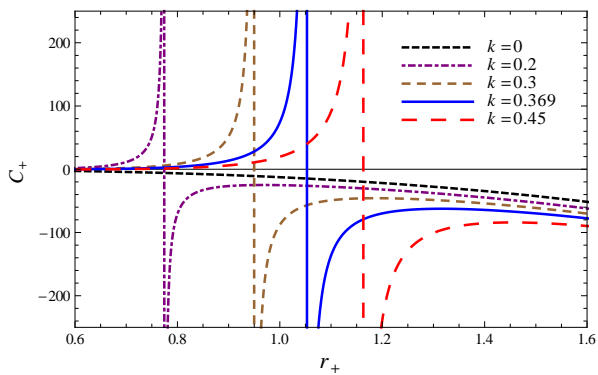}
\end{tabular}
\caption{Plots showing the behaviour of specific heat $C_+$ vs horizon radius $r_+$ for the different values of deviation parameter $k$.}\label{slsf6}
\end{figure*}

 Thus the temperature of the $5D$ regular  black hole is modified due to the deviation parameter $k$. The temperature reduces to $T_+=1/2\pi r_+$ in the absence of deviation parameter $k$. The Fig. \ref{slsf5} demonstrates the behaviour of temperature with radius $r$ for the various values of $k$, and diverges at $k=0$. The expression of the entropy can be obtained by the first law of thermodynamics, 
\begin{equation}
dM_+=T_+\,dS_+,
\end{equation}
one can find the following expression for the black hole entropy
\begin{eqnarray}
S_+=\int\,\frac{1}{T_+}\frac{\partial M_+}{\partial r_+}dr_+= \frac{4}{3\pi r_+^3}\left[\frac{2k+r_+^2}{r_+^2}e^{k/r_+^2}-\frac{8k\sqrt{\pi k}}{r_+^3}\text{erf}\left(\frac{\sqrt{k}}{r_+}\right)\right],
\label{eqS}
\end{eqnarray}
 If we put $k=0$, then we get $S_+=4\pi r_+^3/3$, which is the entropy of $5D$ Schwarzschild-Tangherilini black hole and that entropy is proportional to the area of the event horizon. Next, we study the heat capacity of a black hole, which is essential for study of the thermodynamical stability of black hole \cite{Chaturvedi:2016fea}
\begin{eqnarray}
C_+=\frac{\partial M_+}{\partial T_+}=\Big(\frac{\partial M_+}{\partial r_+}\Big)\Big(\frac{\partial r_+}{\partial T_+}\Big)=-4\pi r_+^3 \left[\frac{r_+^2-k}{r_+^2-3k}\right].
\label{eqC}
\end{eqnarray}
The heat capacity has plotted in Fig.~\ref{slsf6} for different values of $k$. We observed two kinds of behaviour, first one is the positive heat capacity, when $r_+<r_C$, which suggest the thermodynamic stability of the black hole and the other one is the negative heat capacity, when $r_+ > r_C$, which suggest the instability of the black hole \cite{44}. The heat capacity is discontinuous  at $r_+=r_C$ means the second order phase transition occurs.

\begin{thebibliography}{75}

\bibitem{1} A. Einstein, Ann. Phys. (N.Y.), {\bf 49}, 769 (1916); {\bf 14}, 517 (2005).
\bibitem{2}
  W.~de Sitter,
  Mon.\ Not.\ Roy.\ Astron.\ Soc.\  {\bf 78}, 3 (1917);   B.~P.~Abbott {\it et al.} [LIGO Scientific and Virgo Collaborations],
  Phys.\ Rev.\ Lett.\  {\bf 116}, 241102 (2016); Phys.\ Rev.\ Lett.\  {\bf 118}, 221101 (2017); 
  T.~Johannsen, C.~Wang, A.~E.~Broderick, S.~S.~Doeleman, V.~L.~Fish, A.~Loeb and D.~Psaltis,
  Phys.\ Rev.\ Lett.\  {\bf 117}, 091101 (2016).
  
\bibitem{Joshi:2012mk} 
  P.~S.~Joshi and D.~Malafarina,
  Int.\ J.\ Mod.\ Phys.\ D {\bf 20}, 2641 (2011).
  
\bibitem{4} 
  C.~Bambi, D.~Malafarina and L.~Modesto,
  Phys.\ Rev.\ D {\bf 88}, 044009 (2013); 
  L.~Modesto,
  Phys.\ Rev.\ D {\bf 70}, 124009 (2004).
  
\bibitem{Ansoldi:2008jw} 
  S.~Ansoldi,
  arXiv:0802.0330 [gr-qc].

\bibitem{Sakharov:1966aja} 
  A.~D.~Sakharov,
  Sov.\ Phys.\ JETP {\bf 22}, 241 (1966).

\bibitem{Gliner:1966} E. B.~Gliner, Sov.\ Phys.\ JETP {\bf 22}, 378 (1966).

\bibitem{Bardeen:1968} J.~Bardeen, in {\it Proceedings of GR5} (Tiflis, U.S.S.R., 1968).

\bibitem{10}
 T. W. Baumgarte and A. D. Rendall, Class. Quantum Grav. {\bf 10} 327, (1993); 
 R. Brandenberger, V. Mukhanov, and A. Sornborger, Phys. Rev. D {\bf 48}, 1629 (1993).

\bibitem{dym} 
  I.~Dymnikova,
  Gen.\ Rel.\ Grav.\  {\bf 24}, 235 (1992); 
  Class.\ Quant.\ Grav.\  {\bf 19}, 725 (2002);
  Int.\ J.\ Mod.\ Phys.\ D {\bf 12}, 1015 (2003); 
  Class.\ Quant.\ Grav.\  {\bf 21}, 4417 (2004);  
  I.~Dymnikova and E.~Galaktionov,
  Class.\ Quant.\ Grav.\  {\bf 22}, 2331 (2005).

\bibitem{abg}
E. Ayon-Beato and A. Garcia, Phys. Rev. Lett. {\bf 80}, 5056 (1998);  Phys. Lett. B {\bf 464}, 25 (1999);  Gen. Rel. Grav. {\bf 13}, 629 (1999);  Gen. Rel. Grav. {\bf 37}, 635 (2005).

\bibitem{AyonBeato:2000zs} 
  E.~Ayon-Beato and A.~Garcia,
  Phys.\ Lett.\ B {\bf 493}, 149 (2000).
\bibitem{12}
 K. A. Bronnikov,  Phys. Rev. D {\bf 63}, 044005 (2001);  Phys. Rev. D {\bf 64}, 064013 (2001); K. A. Bronnikov, H. Dehnen, and V. N. Melnikov, Phys. Rev. D {\bf 68}, 024025 (2003); Gen. Rel. Grav. {\bf 39}, 973 (2007); K. A. Bronnikov, A. Dobosz, and I. G. Dymnikova,  Class. Quantum Grav. {\bf 20}, 3797 (2003); K. A. Bronnikov and I. Dymnikova, Class.\ Quant.\ Grav.\  {\bf 24}, 5803 (2007);  K. A. Bronnikov and J. C. Fabris,  Phys. Rev. Lett. {\bf 96}, 251101 (2006).

\bibitem{13}
  A.~Borde,
  Phys.\ Rev.\ D {\bf 50}, 3692 (1994); 
  Phys.\ Rev.\ D {\bf 55}, 7615 (1997).
  
\bibitem{Balart:2009xr} 
  L.~Balart,
  Phys.\ Lett.\ B {\bf 687}, 280 (2010).

\bibitem{15}
 Y.~S.~Myung, Y.~W.~Kim and Y.~J.~Park,
  Gen.\ Rel.\ Grav.\  {\bf 41}, 1051 (2009); 
  Phys.\ Lett.\ B {\bf 659}, 832 (2008);
  Phys.\ Lett.\ B {\bf 656}, 221 (2007); 
  JHEP {\bf 0702}, 012 (2007).
  
  \bibitem{16}
  M.~Amir, F.~Ahmed and S.~G.~Ghosh,
  Eur.\ Phys.\ J.\ C {\bf 76}, 532 (2016); 
  S.~G.~Ghosh and M.~Amir,
  Eur.\ Phys.\ J.\ C {\bf 75}, 553 (2015); 
  S.~G.~Ghosh, P.~Sheoran and M.~Amir,
  Phys.\ Rev.\ D {\bf 90}, 103006 (2014); 
  F.~Ahmed, M.~Amir and S.~G.~Ghosh,
  arXiv:1805.00804 [gr-qc].

\bibitem{Lemos:2011dq} 
  J.~P.~S.~Lemos and V.~T.~Zanchin,
  Phys.\ Rev.\ D {\bf 83}, 124005 (2011).
  
\bibitem{Hayward:2005gi} 
  S.~A.~Hayward,
  Phys.\ Rev.\ Lett.\  {\bf 96}, 031103 (2006).

\bibitem{19}
  L.~Balart and E.~C.~Vagenas,
  Phys.\ Lett.\ B {\bf 730}, 14 (2014); 
  Z.~Y.~Fan and X.~Wang,
  Phys.\ Rev.\ D {\bf 94}, 124027 (2016);  
D.~V.~Singh, M.~S.~Ali and S.~G.~Ghosh, Int. J. Mod. Phys. D {\bf 27}, 1850108 (2018).

\bibitem{Xiang:2013sza} 
  L.~Xiang, Y.~Ling and Y.~G.~Shen,
  Int.\ J.\ Mod.\ Phys.\ D {\bf 22}, 1342016 (2013);
  H.~Culetu,
  Int.\ J.\ Theor.\ Phys.\  {\bf 54}, 2855 (2015).
  
  \bibitem{Toshmatov} 
 B.~Toshmatov, B.~Ahmedov, A.~Abdujabbarov and Z.~Stuchlik,
  Phys.\ Rev.\ D {\bf 89}, 104017 (2014);

\bibitem{Bambi:2013ufa} 
  C.~Bambi and L.~Modesto,
  Phys.\ Lett.\ B {\bf 721}, 329 (2013).

\bibitem{Neves:2014aba} 
  J.~C.~S.~Neves and A.~Saa,
  Phys.\ Lett.\ B {\bf 734}, 44 (2014).

\bibitem{Larranaga:2014uca} 
  A.~Larranaga, A.~Cardenas-Avendano and D.~A.~Torres,
  Phys.\ Lett.\ B {\bf 743}, 492 (2015).
  
\bibitem{Ghosh:2014pba} 
  S.~G.~Ghosh,
  Eur.\ Phys.\ J.\ C {\bf 75}, 532 (2015).
  
\bibitem{Newman:1965tw} 
  E.~T.~Newman and A.~I.~Janis,
  J.\ Math.\ Phys.\  {\bf 6}, 915 (1965).
  
\bibitem{27} 
  M.~Azreg-Aïnou,
  Phys.\ Rev.\ D {\bf 90}, 064041 (2014); 
  Phys.\ Lett.\ B {\bf 730}, 95 (2014); 
  Eur.\ Phys.\ J.\ C {\bf 74}, 2865 (2014).

\bibitem{kerr} R. P. Kerr, Phys. Rev. Lett. {\bf 11}, 237 (1963); H.~Erbin and L.~Heurtier,
  Class.\ Quant.\ Grav.\  {\bf 32}, 165004 (2015).


\bibitem{Ghosh:2018bxg} 
  S.~G.~Ghosh, D.~V.~Singh and S.~D.~Maharaj,
  Phys.\ Rev.\ D {\bf 97}, 104050 (2018).

\bibitem{Hendi:2017phi} 
  S.~H.~Hendi, N.~Riazi, S.~Panahiyan and B.~Eslam Panah,
  arXiv:1710.01818 [gr-qc].
  
\bibitem{Panahiyan:2018gzq} 
  S.~Panahiyan, S.~H.~Hendi and N.~Riazi,
  arXiv:1802.00309 [gr-qc].

\bibitem{Rizzo:2006zb} 
  T.~G.~Rizzo,
  JHEP {\bf 0609}, 021 (2006).

\bibitem{Wheeler}
 C. W. Misner, K.S. Thorne and J.A. Wheeler, {\it Gravitation}, (W. H. Freeman, San Francisco,
1973).

\bibitem{Singh:2017qur} 
  D.~V.~Singh and N.~K.~Singh,
  Annals Phys.\  {\bf 383}, 600 (2017);   D.~V.~Singh, S.~G.~Ghosh and S.~D.~Maharaj,
  Annals Phys.\  {\bf 412} (2020) 168025;   A.~Kumar, D.~Veer Singh and S.~G.~Ghosh,
  Eur.\ Phys.\ J.\ C {\bf 79} (2019)  275;  D.~V.~Singh and S.~Siwach,
  arXiv:1909.11529 [hep-th].

\bibitem{Penrose:1971uk} 
  R.~Penrose and R.M.~Floyd,
  Nature {\bf 229}, 177 (1971).
  
\bibitem{Papnoi:2014aaa} 
  U.~Papnoi, F.~Atamurotov, S.~G.~Ghosh and B.~Ahmedov,
  Phys.\ Rev.\ D {\bf 90}, 024073 (2014).

\bibitem{Johannsen:2015qca} 
  T.~Johannsen,
  Astrophys.\ J.\  {\bf 777}, 170 (2013).

\bibitem{Amir:2017slq} 
  M.~Amir, B.~P.~Singh and S.~G.~Ghosh,
  Eur.\ Phys.\ J.\ C {\bf 78}, 399 (2018).

\bibitem{Hioki:2009na} 
  K.~Hioki and K.~I.~Maeda,
  Phys.\ Rev.\ D {\bf 80}, 024042 (2009).
  
\bibitem{Amir:2016cen} 
  M.~Amir and S.~G.~Ghosh,
  Phys.\ Rev.\ D {\bf 94}, 024054 (2016). 

\bibitem{41}
  S.~R.~Das, G.~W.~Gibbons and S.~D.~Mathur,
  Phys.\ Rev.\ Lett.\  {\bf 78}, 417 (1997);  
  Y.~Decanini, G.~Esposito-Farese and A.~Folacci,
  Phys.\ Rev.\ D {\bf 83}, 044032 (2011).

\bibitem{Mashhoon:1973zz} 
  B.~Mashhoon,
  Phys.\ Rev.\ D {\bf 7}, 2807 (1973).

\bibitem{Chaturvedi:2016fea} 
  P.~Chaturvedi, N.~K.~Singh and D.~V.~Singh,
  Int.\ J.\ Mod.\ Phys.\ D {\bf 26}, 1750082 (2017).

\bibitem{44}
  S.~W.~Hawking and D.~N.~Page,
  Commun.\ Math.\ Phys.\  {\bf 87}, 577 (1983); 
  P.~C.~W.~Davies,
  Proc.\ Roy.\ Soc.\ Lond.\ A {\bf 353}, 499 (1977).

  
  
  \end{thebibliography}
\end{document}